\documentclass[12pt]{article}
\usepackage[table,xcdraw]{xcolor}
\usepackage{amsmath}
\usepackage{amsmath,bm}
\usepackage{url}
\usepackage{tikz}
\usetikzlibrary{shapes.geometric}
\usetikzlibrary{arrows}
\usepackage{amsmath}
\usepackage{amssymb}
\usepackage{hyperref}
\usepackage{rotating}
\usepackage{physics}
\usepackage{epsfig}
\usepackage{graphicx}
\usepackage{color}
\usepackage{cite}
\usepackage{subfig}
\usepackage{slashed}
\usepackage[font=small,labelfont=bf]{caption}

\newcommand{\beq}{\begin{equation}}
\newcommand{\eeq}{\end{equation}}
\newcommand{\ga}{\lower.7ex\hbox{$\;\stackrel{\textstyle>}{\sim}\;$}}
\newcommand{\la}{\lower.7ex\hbox{$\;\stackrel{\textstyle<}{\sim}\;$}}

\hypersetup{
    colorlinks = true,
    citecolor = {blue},
    linkcolor = {blue},
    urlcolor = {blue},
}

\begin{document}

\def\jcap{\ref@jnl{J. Cosmology Astropart. Phys.}}

\begin{flushright}
{\tt KCL-PH-TH/2020-14}, {\tt CERN-TH-2020-046}  \\
{\tt ACT-02-20, MI-TH-2010} \\
{\tt UMN-TH-3915/20, FTPI-MINN-20/05} \\
\end{flushright}

\begin{center}
{\bf {\large Phenomenology and Cosmology of No-Scale Attractor \\[0.2cm] 
Models of Inflation}}

\end{center}

\begin{center}{
{\bf John~Ellis}$^{a}$,
{\bf Dimitri~V.~Nanopoulos}$^{b}$,
{\bf Keith~A.~Olive}$^{c}$ and
{\bf Sarunas~Verner}$^{c}$}
\end{center}

\begin{center}
{\em $^a$Theoretical Particle Physics and Cosmology Group, Department of
  Physics, King's~College~London, London WC2R 2LS, United Kingdom;\\
Theoretical Physics Department, CERN, CH-1211 Geneva 23,
  Switzerland;\\
  National Institute of Chemical Physics \& Biophysics, R{\" a}vala 10, 10143 Tallinn, Estonia}\\[0.3cm]
{\em $^b$George P. and Cynthia W. Mitchell Institute for Fundamental
 Physics and Astronomy, Texas A\&M University, College Station, TX
 77843, USA};\\
{\em Astroparticle Physics Group, Houston Advanced Research Center (HARC),
 \\ Mitchell Campus, Woodlands, TX 77381, USA;\\ 
Academy of Athens, Division of Natural Sciences,
Athens 10679, Greece}\\[0.3cm]
{\em $^c$William I. Fine Theoretical Physics Institute, School of
 Physics and Astronomy, University of Minnesota, Minneapolis, MN 55455,
 USA}
 
 \end{center}

\vspace{0.3cm}
\centerline{\bf ABSTRACT}
\vspace{0.1cm}

{\small We have recently proposed attractor models for modulus fixing, inflation, supersymmetry breaking and dark energy based on no-scale supergravity. In this paper we develop phenomenological and cosmological aspects of these no-scale attractor models that underpin their physical applications. We consider models in which inflation is driven by a modulus field ($T$-type) with supersymmetry broken by a Polonyi field, or a matter field ($\phi$-type) with supersymmetry broken by the modulus field. We derive the possible patterns of soft supersymmetry-breaking terms, which depend in $T$-type models whether the Polonyi and/or matter fields are twisted or not, and in $\phi$-type models on whether the inflaton and/or other matter fields are twisted or not. In $\phi$-type models, we are able to directly relate the scale of supersymmetry breaking to the inflaton mass.  We also discuss cosmological constraints from entropy considerations and the density of dark matter on the mechanism for stabilizing the  modulus field via higher-order terms in the no-scale K\"ahler potential.
 } 

\vspace{0.2in}

\begin{flushleft}
{April} 2020
\end{flushleft}
\medskip
\noindent

\newpage

\section{Introduction}
The theory of inflation is the most successful phenomenological framework for describing the near-flatness of the Universe, as well as explaining why it appears to be statistically homogeneous and isotropic on large scales~\cite{reviews}. The current measurements of the cosmic microwave background~(CMB) from the Planck satellite are perfectly consistent with the inflationary paradigm. They exhibit an almost scale-invariant spectrum of scalar perturbations with tilt $n_s \simeq 0.96$ to $0.97$ \cite{planck18} and no discernible non-Gaussianities, with an upper limit of tensor-to-scalar ratio $r \lesssim 0.06$ \cite{rlimit}. The combination of cosmological observables $n_s$ and $r$ already discriminates between different models of inflation, excluding simple monomial scalar potentials whilst being consistent with the Starobinsky model~\cite{Staro}, and future CMB measurements will further constrain the surviving models of inflation.

In order to connect inflation to a viable quantum theory of gravity at high scales and to the Standard Model (SM) of particle physics at lower scales, we are motivated to consider models based on no-scale supergravity \cite{no-scale,Ellis:1983sf,LN}. It was shown in \cite{Witten} that no-scale supergravity appears generically in string theory compactifications, which we regard as the UV completion of no-scale supergravity models, and it has also been shown how the SM can be incorporated in no-scale models of inflation~\cite{ENO8,EGNO4,king2,egnno1,EGNNO23,EGNNO45,dgmo}. 

The original Starobinsky model of inflation \cite{Staro}, which is based on $R + R^2$ gravity, leads to a scalar tilt value of \cite{MukhChib} $n_s \simeq 0.965$ and a scalar-to-tensor ratio $r \simeq 0.003$, and is entirely consistent with the current CMB data. It was shown in \cite{ENO6} that one can easily obtain a Starobinsky-like potential in the context of no-scale supergravity. Moreover, it was further shown that there are many
inflationary avatars within the no-scale framework~\cite{Avatars}, including the no-scale attractor models that we discuss in this paper, and we provided a general classification of these models in~\cite{ENOV1}. We extended these models in~\cite{ENOV2,ENOV3}, and combined supersymmetry breaking and dark energy with a Starobinsky-like model of inflation.

This latter point is crucial, as constructing models with
acceptable phenomenology, cosmology and supersymmetry breaking has been notoriously difficult, particularly when combined with models of inflation that are consistent with the CMB data. Although first steps were made in~\cite{ENOV2}, many more detailed issues remain to be studied. The main goal of this paper is to develop further the phenomenology and cosmology of no-scale attractor models, bridging the gap between string inspiration and a viable scenario incorporating dark matter and the SM.

In particular, we show how to construct various successful no-scale attractor models of inflation, characterize different possibilities for supersymmetry breaking, and discuss cosmology following inflation and the constraints imposed by entropy considerations and the dark matter density on mechanisms for field stabilization via higher-order terms in the K\"ahler potential. 

We distinguish two types of no-scale attractor models: in the first type the inflaton is identified with a volume modulus field, denoted by $T$, and in the second type the inflaton is identified with a matter field, denoted by $\phi$. In $T$-type models 
supersymmetry is broken in a hidden Polonyi sector~\cite{pol}, whereas in $\phi$-type models supersymmetry is broken by the $T$ field, in which case there is no need for an additional sector to break supersymmetry.
Supersymmetry breaking in the $T$-type models was studied in~\cite{strongmoduli, EGNO4}, and their crucial feature is that they favour boundary conditions with universal soft scalar masses, as in minimal supergravity (mSUGRA) models. The $\phi$-type models were discussed in~\cite{EGNO4,ENOV2,ENOV3}, and they open up various less constrained phenomenological possibilities, including sources for non-universal scalar masses, as we discuss in this paper. In principle, these different boundary conditions for supersymmetry breaking have distinctive phenomenological and cosmological features that may be used to distinguish between models. We show that in $\phi$-type models it is possible to relate the scale of supersymmetry breaking to the inflationary scale without fine-tuning. 

The structure of this paper is as follows. In Section~2 we review general features of no-scale supergravity, and introduce the general category of no-scale attractor models (some details are in Appendix~A). Then, in Section~3 we discuss $T$-type attractor scenarios, introducing the needed Polonyi sector, which may be either twisted or untwisted, discussing the possibilities for either twisted or untwisted matter fields, and presenting the corresponding predictions for soft supersymmetry-breaking parameters (some details are in Appendix~B). Section~4 contains an analogous discussion of $\phi$-type attractor scenarios in which either the inflaton and/or the other matter fields may be either twisted or not. As we discuss in Section~5, both classes of models require the stabilization of some field, the Polonyi field or the modulus field, respectively. We discuss the post-inflationary dynamics in the various cases and the corresponding constraints due to entropy considerations and the dark matter density. Finally, Section~6 presents some conclusions and prospects.

\section{No-Scale Supergravity and Inflation}

We begin our discussion by recalling some general features of no-scale inflationary models. Minimal $\mathcal{N} = 1$ no-scale supergravity models were first discussed in~\cite{no-scale,Ellis:1983sf, EKN1, LN}, and are characterized by the following K\"ahler potential form with a single chiral field $T$,
\begin{equation}
\label{kah1}
K \; = \; -3 \, \ln(T + \overline{T}),
\end{equation}
which parameterizes a non-compact $\frac{SU(1, 1)}{U(1)}$ coset K\"ahler manifold, whose scalar curvature is given by $R = \frac{2}{3}$.~\footnote{We choose the convention where $R < 0$ corresponds to a spherical manifold and $R > 0$ corresponds to a hyperbolic manifold.} 

In order to construct no-scale attractor models~\cite{ENOV3}, we consider the following
generalization~\cite{EKN1} of the K\"ahler potential~(\ref{kah1}):
\begin{equation}
\label{kah2}
K \; = \; -3 \, \alpha \ln(T + \overline{T}),
\end{equation}
which incorporates a free positive curvature parameter $\alpha$ (see also~\cite{EKN1,alpha1, alpha3}). In this case, the K\"ahler potential form~(\ref{kah2}) still parameterizes an $\frac{SU(1, 1)}{U(1)}$ coset K\"ahler manifold, but with scalar curvature given by $R = \frac{2}{3 \alpha}$.  

To construct realistic no-scale attractor models of inflation, we need to extend the K\"ahler potential form~(\ref{kah2}). One minimal possibility is to introduce an additional `untwisted' matter-like chiral field $\phi$:
\begin{equation}
\label{kah3}
K \; = \; -3 \, \alpha \ln(T + \overline{T} - \frac{|\phi|^2}{3}),
\end{equation}
which characterizes a non-minimal $\frac{SU(2,1)}{SU(2) \times U(1)} $ K\"ahler manifold. One can also consider models with a `twisted' matter-like field $\varphi$ via the K\"ahler potential:
\begin{equation}
\label{kah4}
K \; = \; -3 \, \alpha \ln(T + \overline{T}) + |\varphi|^2,
\end{equation}
which parameterizes a non-minimal $\frac{SU(1,1)}{U(1)} \times U(1)$ K\"ahler manifold.

To extend the model beyond inflation and include low energy phenomenological interactions, the K\"ahler potential must include additional fields to account for Standard Model (SM) particles~\cite{EKN2}, e.g.,
\begin{equation}
\label{kah3i}
K \; = \; -3 \, \alpha \ln(T + \overline{T} - \sum_i \frac{ |\phi_i|^2}{3})
\end{equation}
in the untwisted case, which parameterizes a $\frac{SU(N,1)}{SU(N) \times U(1)} $ K\"ahler manifold.
It was shown in~\cite{Witten} that the K\"ahler potential form~(\ref{kah3i}) with $\alpha = 1$ emerges as the low-energy effective theory of one simple string compactification scenario in which the complex scalar field $T$ corresponds to the compactification volume modulus, whereas other values of $\alpha$ can be found in other scenarios.
Similarly, one can consider models in which the extra matter fields are twisted, or a mixture of twisted and untwisted fields. 

In order to accommodate SM interactions, we include a superpotential $W$ appearing via the extended K\"ahler potential
\begin{equation}
G \; = \; K + \ln~W + \ln~\overline{W},
\end{equation}
which then leads to the following expression for the effective scalar potential,
\begin{equation}
\label{effpot}
V \; = \; e^{G} \left[\pdv{G}{\Phi_i} \left( K^{-1} \right)_{i}^{j} \pdv{G}{\bar{\Phi}^{j}} - 3 \right],
\end{equation}
where the $\Phi_i$ are complex chiral fields, the $\bar{\Phi}_{\bar{j}}$ are their conjugate fields, and $K^{i}_j \equiv \partial^2 K/ \partial \Phi_i \partial \bar{\Phi}^{j}$ is the K\"ahler metric.

As already mentioned, we discuss in this paper two different types of no-scale attractor models of inflation: models in which the inflaton field is identified with the volume modulus, $T$, and supersymmetry is broken by introducing~\cite{pol} a Polonyi field, $Z$, and models where the inflaton field is identified with a matter-like field, $\phi$, and supersymmetry can be broken without invoking a hidden Polonyi sector. We refer to the former as $T$-type models and to the latter as $\phi$-type models.

We focus primarily on $\alpha$-Starobinsky models of inflation based on a scalar potential given by ~\footnote{We work in units of the reduced Planck mass, $M_P = \frac{1}{\sqrt{8 \pi G}} \simeq 2.4 \times 10^{18} \, \text{GeV}$. In most cases, factors of $M_P$ are omitted, particularly for fields, their expectation values and stabilization terms.}
\begin{equation}
V = \frac{3}{4} M^2 \left(1 - e^{- \sqrt{\frac{2}{3 \alpha}} x} \right)^{2},
\label{astaro}
\end{equation}
where $x$ is a canonically-normalized inflaton field and $M$ is the inflaton mass. In such an $\alpha$-Starobinsky model, the cosmological observables~$(n_s, r)$ (where $n_s$ is the tilt in the scalar perturbation spectrum and $r$ is the tensor-to-scalar ratio) depend on the K\"ahler curvature parameter $\alpha$. A first discussion of such models was presented in~\cite{Avatars}, where it was shown that, for $\alpha \lesssim \mathcal{O} (1)$, $\alpha$-Starobinsky models of inflation predict
\begin{equation}
\label{cmb}
n_s \simeq 1 -\frac{2}{N_{*}} \, , \qquad r \simeq \frac{12 \alpha}{N_{*}^2} \, ,
\end{equation}
where $N_{*}$ is the number of e-folds of inflation.
Eliminating $N_*$ from~(\ref{cmb}),
the curvature parameter can be expressed as follows in terms of the cosmological observables:
\begin{equation}
\label{observables}
\alpha \simeq \frac{r}{3 \left(1 - n_s \right)^2}.
\end{equation}
for $\alpha \lesssim \mathcal{O} (1)$.

It is possible to find analytic formulae for $n_s$ and $r$ that hold without any restriction on $\alpha$, but they involve special functions and are given in Appendix~A. We show in Fig.~\ref{one} curves of $n_s$ and $r$ as functions of $\alpha$ for $N_* = 50, 55$, and $60$. The current observation range\footnote{We are applying Planck results \cite{planck18} based on the combination of TT,TE,EE+lowE+lensing data for $n_s$ and the combination of BICEP2, Keck Array, and Planck data \cite{rlimit} for $r$.} $n_s \in [0.961, 0.969]$ does not constrain $\alpha$ significantly. However, the current Planck upper limit $r < 0.06$ imposes the upper limit $\alpha \lesssim 46 ~(88)$ for $N_* = 50 ~(60)$. Future CMB observations~\cite{LiteBIRD} should be able to probe the tensor-to-scalar ratio down to an upper limit $r  \lesssim 0.001$, which is sufficient to determine accurately the curvature parameter $\alpha$, with a lower limit $\alpha \gtrsim 0.3$ when $N_* \simeq 60$, and underpin cosmological string phenomenology. 
In the following Sections we discuss the distinctive phenomenological features and cosmological aspects of both twisted and untwisted models.

\begin{figure}[h!]
\centering
\includegraphics[scale=0.395]{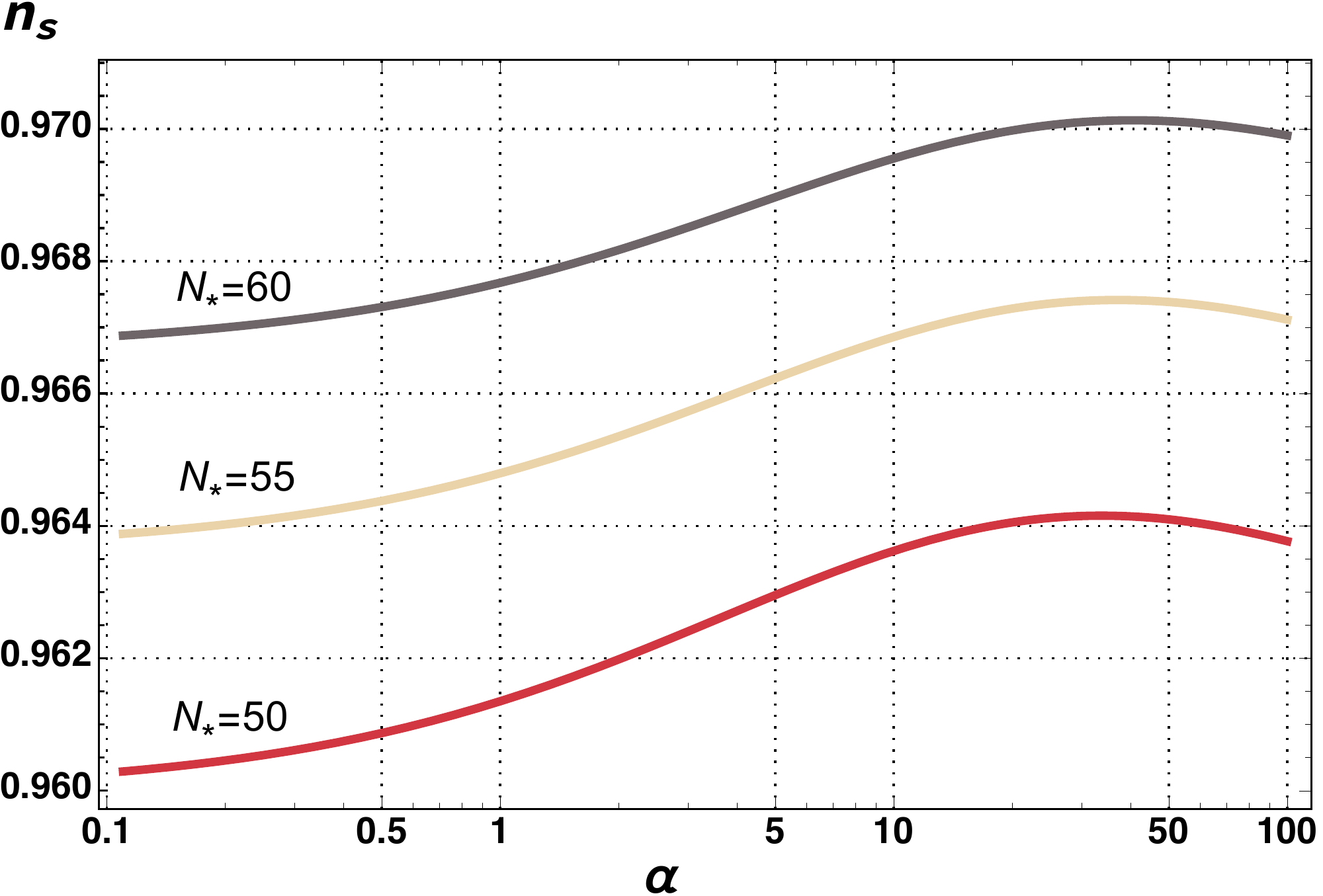}
\includegraphics[scale=0.395]{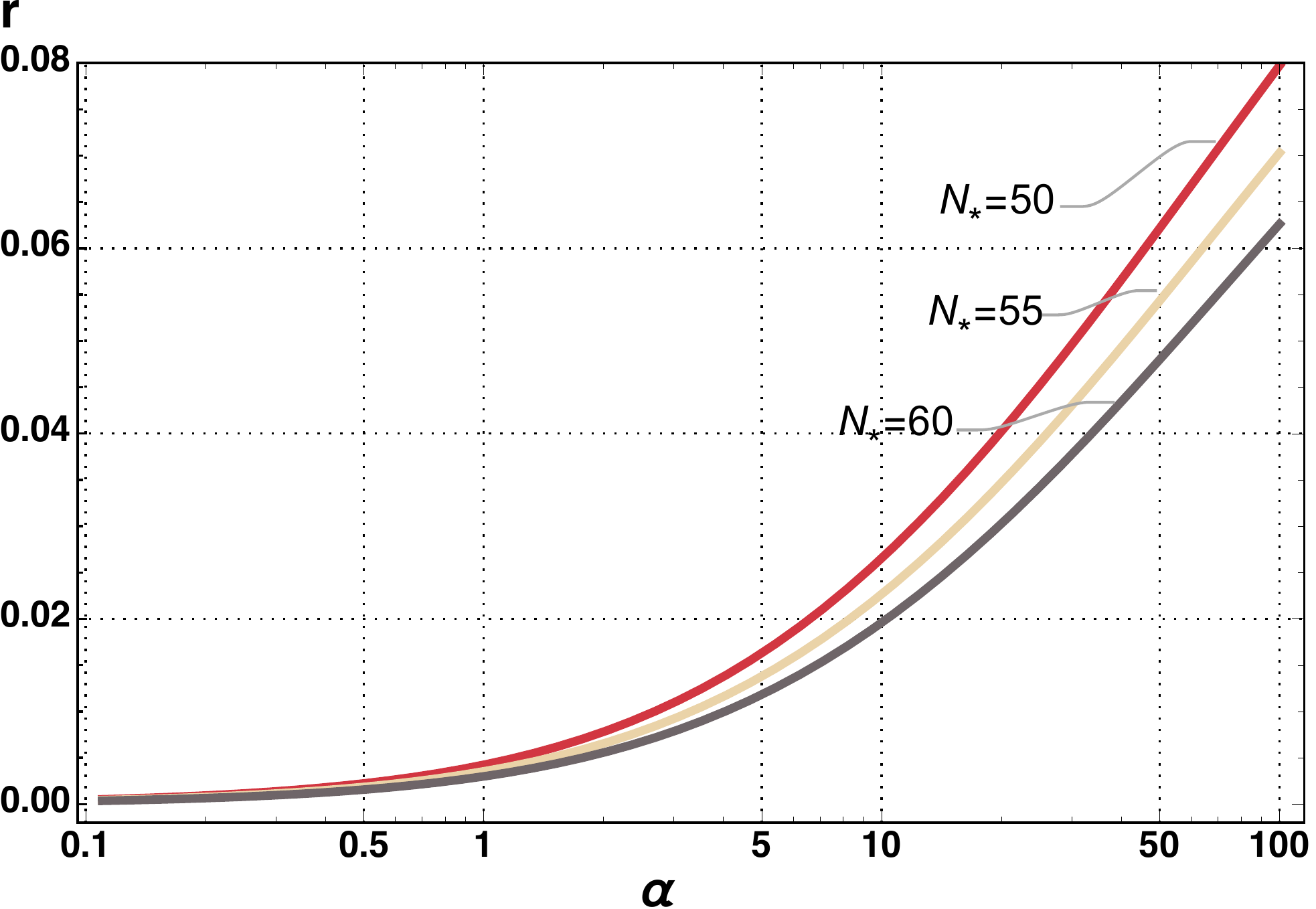}\\
\caption{\it Plots of $n_s$ (left panel) and $r$ (right panel) as functions of $\alpha$ for the representative values $N_* = 50, 55, 60$.}
\label{one}
\end{figure}

\section{$T$-type No-Scale Attractor Scenarios}
\subsection{General Framework}
We begin with general no-scale attractors based on the non-minimal $\frac{SU(2,1)}{SU(2) \times U(1)}$ K\"ahler potential~(\ref{kah3}), where the volume modulus $T$ drives inflation. We consider the following general form of inflationary superpotential \cite{klr}:
\begin{equation}
\label{C1}
W_I = \sqrt{\alpha} \, M \phi \, f(T) \,\left(2T \right)^{\frac{3 \alpha - 1}{2}},
\end{equation}
where $f(T)$ is an arbitrary function of the volume modulus $T$ only, and $M$ is the inflaton mass scale. 
For $\alpha = 1$, this reduces to the supergravity version of the $R + R^2$ model discussed in \cite{Cecotti,KLno-scale,Avatars,EGNO123,rs,klr,reheating,moreothers}.  If we combine the superpotential~(\ref{C1}) with the K\"ahler potential~(\ref{kah3}), the effective scalar potential~(\ref{effpot}) in the real $T$ direction becomes:
\begin{equation}
\label{potT1}
V = M^2 \, f(T)^2,
\end{equation}
where we have assumed that the vacuum expectation value (VEV) of the matter-like field $\phi$ is fixed to $\langle \phi \rangle = 0$, which can achieved by introducing higher-order stabilization terms in the K\"ahler potential~(\ref{kah3}), as we discuss below \cite{Avatars,EKN3}.

It was discussed in~\cite{ENOV3} that one can obtain Starobinsky-like models of inflation from such a superpotential (\ref{C1}) when the volume modulus $T$ is associated with the inflaton. The $\alpha$-Starobinsky model can be obtained by considering the function:
\begin{equation}
\label{cecfunc}
 f(T) =  \frac{\sqrt{3}}{2T} \left(T - \frac{1}{2} \right),
\end{equation}
in which case the effective potential~(\ref{potT1}) becomes
\begin{equation}
\label{pot11}
V = \frac{3 M^2}{4T^2} \left(T - \frac{1}{2} \right)^2.
\end{equation}
Defining a canonically-normalized field $\rho \equiv \sqrt{\frac{3 \alpha}{2}} \ln 2T$, the scalar potential~(\ref{pot11}) can be rewritten in the $\alpha$-Starobinsky inflationary form
given in Eq.~(\ref{astaro})
with $x=\rho$ driving inflation.

More generally, this framework can be applied to any form of effective scalar potential that vanishes when the volume modulus $T$ obtains a vacuum expectation value, as long as $f(\langle T \rangle) = 0$.~\footnote{The representative example above has 
$\langle T \rangle = 1/2$, but this choice is arbitrary and models with other values 
of $\langle T \rangle$ yield similar results.}.

\subsection{Supersymmetry Breaking with a Polonyi Field}
\label{sec:TZ}

In this Section we discuss possible patterns of supersymmetry breaking in $T$-type models, which we
accomplish by introducing a Polonyi field $Z$ \cite{pol} with a non-vanishing $F$-term.~\footnote{Introducing a constant term in the superpotential~(\ref{C1}) would shift the potential minimum to a supersymmetry-preserving AdS minimum \cite{EGNO4}, rather than break supersymmetry.}

We can consider the K\"ahler potential~(\ref{kah3}) with either
an untwisted and strongly-stabilized Polonyi field \cite{strongpol,Dudas:2006gr,klor,dlmmo,eioy,nataya,ADinf,strongmoduli,EGNO4,dgmo}, given by
\begin{equation}
\label{kahtuntwist1}
K \; = \; -3 \, \alpha \ln \left( T + \overline{T} - \frac{|\phi|^2}{3} + \frac{|\phi|^4}{\Lambda_{\phi}^2} - \frac{|Z|^2}{3}  + \frac{|Z|^4}{\Lambda_Z^2} \right),
\end{equation}
or a twisted and strongly-stabilized Polonyi field,
\begin{equation}
\label{kahttwist1}
K \; = \; -3 \, \alpha \ln \left( T + \overline{T} - \frac{|\phi|^2}{3} + \frac{|\phi|^4}{\Lambda_{\phi}^2} \right) + |Z|^2 - \frac{|Z|^4}{\Lambda_Z^2} ,
\end{equation}	
where we have also introduced a quartic stabilization term for the matter-like field $\phi$, which fixes dynamically its VEV to $\langle \phi \rangle = 0$ during inflation \cite{EKN3,Avatars}. We can consider a general form of the function $f(T)$ (\ref{C1}) with $f(\langle T \rangle) = f(1/2) = 0$, which we express as $f(T) = c(T)\left(T-\frac{1}{2}\right)$, where $c(\langle T \rangle) \equiv c$ and $f'(\langle T \rangle) = c$. For example, the superpotential~(\ref{cecfunc}) with $\langle T \rangle = 1/2$ gives $c = \sqrt{3}$.	

Next, we introduce the following Polonyi superpotential \cite{pol}
\begin{align}
\label{pol}
 W_P \; =\mu \left(Z + b \right),
\end{align}
 which is responsible for supersymmetry breaking, and
 $b$ is a constant. In the absence of strong stabilization,
minimization of the Polonyi potential at zero vacuum energy leads to the solution that $\langle z \rangle = \sqrt{3} - 1$ and $b=2-\sqrt{3}$. In models with a strongly-stabilized Polonyi field, the minimum of the potential with zero vacuum energy is near the origin and $\langle z \rangle \propto \Lambda_Z^2$ with $b = 1/\sqrt{3}$ (for $\alpha = 1$) \cite{ADinf,strongmoduli}.   If we consider the combined superpotential $W_I + W_P$, where $W_I$ is given by~(\ref{C1}), the minimum of the effective scalar potential shifts \cite{EGNO4,dgmo} and we find new VEVs for our fields. The shifted VEVs for K\"ahler potentials~(\ref{kahtuntwist1}) and~(\ref{kahttwist1}) are 
given by
\begin{equation}
\label{shiftone}
\begin{aligned}[c]
& {\rm Untwisted~Case}: \\
& \langle T \rangle \simeq \frac{1}{2} + \left(\frac{2 \alpha - 1}{ \alpha c^2}\right) \Delta^2,\\
& \langle \phi \rangle \simeq \frac{\sqrt{3}}{c} \Delta,\\
& \langle Z \rangle \simeq \frac{\sqrt{\alpha}}{6 \sqrt{3}} \Lambda_Z^2, \\
& b  \simeq \frac{1}{\sqrt{3 \alpha}} - \left( \frac{1 + 3 \alpha(\alpha - 1)}{2 \sqrt{3} \alpha^{3/2} c^2} \right) \Delta^2,
\end{aligned}
\qquad \qquad
\begin{aligned}[c]
& {\rm Twisted~Case}: \\
& \langle T \rangle \simeq \frac{1}{2} + \frac{2 \alpha}{c^2} \Delta^2,\\
& \langle \phi \rangle \simeq \frac{\sqrt{3 \alpha}}{c} \Delta,\\
& \langle Z \rangle \simeq \frac{1}{2 \sqrt{3}} \Lambda_Z^2, \\
& b  \simeq \frac{1}{\sqrt{3}} - \frac{\sqrt{3} \alpha^2}{2 c^2} \Delta^2,
\end{aligned}
\end{equation}
where we define $\Delta \equiv \mu/M$ 	and assume that $\Delta, \Lambda_Z \ll 1$. It is important to note that the VEVs of the shifted fields and the induced soft parameters depend on the curvature parameter $\alpha$ in the K\"ahler potentials~(\ref{kahtuntwist1}) and~(\ref{kahttwist1}). If we consider the original model~\cite{Cecotti} with the choice $\alpha = 1$, we recover the results in~\cite{EGNO4,dgmo}.
 
As mentioned at the beginning of this Section, supersymmetry is broken through a non-vanishing $F$-term for the Polonyi field $Z$, which is given by
\begin{align}
\label{ftermunt}
   {\rm Untwisted~Case:} & \qquad \sum_{i=1}^{3} |F_i|^2  =  |e^{G/2} \left( K^{-1} \right)_i^j G_j|^2 \simeq |F_Z|^2 \simeq \frac{\mu^2}{\alpha} \simeq 3 m_{3/2}^2\, , \\
   \label{ftermtw}
   {\rm Twisted~Case:} & \qquad \sum_{i=1}^{3} |F_i|^2  =  |e^{G/2} \left( K^{-1} \right)_i^j G_j|^2 \simeq |F_Z|^2 \simeq \mu^2 \simeq 3 m_{3/2}^2\, .
\end{align}
  where the gravitino mass $m_{3/2}$ is given simply by
\begin{align}
\label{gravunt}
   {\rm Untwisted~Case:} & \qquad m_{3/2} \simeq \frac{\mu}{\sqrt{3 \alpha}}\, , \\
\label{gravtw}   
    {\rm Twisted~Case:} & \qquad m_{3/2} \simeq \frac{\mu}{\sqrt{3}}\, .
    \end{align}
Further, we introduce a canonical parameterization of the complex Polonyi field $Z$:
\begin{align}
\label{zcanpar1}
   {\rm Untwisted~Case:}&  \qquad Z = \sqrt{3} \tanh(\frac{z}{\sqrt{6 \alpha}})\, , \\
\label{zcanpar2}
   {\rm Twisted~Case:}& \qquad Z = \frac{z}{\sqrt{2}}\, ,
\end{align}    
and we assume that the imaginary component of the complex field $Z$ vanishes, which is achieved dynamically with the help of a stabilization term parameterized by $\Lambda_Z$.
The mass of the canonically-normalized Polonyi field $z$ is then given by
\begin{equation}
\label{pol1}
{\rm Untwisted~Case:} \qquad m_z^2 \simeq \frac{36 m_{3/2}^2}{\Lambda_Z^2}\, ,
\end{equation}
\begin{equation}
\label{pol2}
{\rm Twisted~Case:} \qquad m_z^2 \simeq \frac{12 m_{3/2}^2}{ \Lambda_Z^2}\, ,
\end{equation}
which is heavier than the gravitino mass $m_{3/2}$ in both cases when $\Lambda_Z \lesssim {\cal O}(1)$. This mass hierarchy between $z$ and the gravitino is instrumental in alleviating \cite{strongmoduli} the so-called cosmological moduli problem \cite{polprob}. Using the field VEVs~(\ref{shiftone}), we can express the Goldstino field as
\begin{equation}
{\rm Untwisted~Case:} \qquad    \eta =\sum_{i = 1}^3 G^i \chi_i \simeq \sqrt{3 \alpha} \chi_z \, ,
\end{equation}
\begin{equation}
{\rm Twisted~Case:} \qquad     \eta = \sum_{i = 1}^3 G^i \chi_i \simeq \sqrt{3 } \chi_z \, ,
\end{equation}
where we see that the Goldstino is the fermionic partner of supersymmetry-breaking Polonyi field $Z$, as expected.

\subsection{Incorporation of Matter Particles}
\label{sec:TZX}

We are now in a position to extend the model to include a general superpotential form that incorporates matter-like fields $X_i$ such as appear in the SM:
\begin{equation}
W = W_{I}(T, \phi) + W_P(Z) + W_2 (X_i) + W_3 (X_i),
\label{WType}
\end{equation}
where $W_I$ is our inflationary superpotential and we have introduced general bilinear and trilinear couplings $W_{2,3}$. The kinetic terms for the matter fields may originate
as untwisted or twisted fields. In the case
of untwisted matter fields, their contributions to the K\"ahler potential lies inside the logarithmic term in either Eqs.~(\ref{kahtuntwist1}) or~(\ref{kahttwist1}):
\begin{equation}
\label{untwisted}
{\rm Untwisted~Matter~Fields:} \qquad    K \supset - \frac{|X_i|^2}{3} .
\end{equation}
For twisted matter fields, the contribution to K 
\begin{equation}
\label{twisted}
{\rm Twisted~Matter~Fields:} \qquad    K \supset |X_i|^2,
\end{equation}
sits outside the logarithmic terms in  Eqs.~(\ref{kahtuntwist1}) and~(\ref{kahttwist1}). 

Having introduced matter fields (twisted and/or untwisted) and supersymmetry breaking via a Polonyi sector, we are now in a position to calculate the soft supersymmetry breaking terms for each of the four possible cases. 
In each case, the soft supersymmetry-breaking terms in 
the Lagrangian are written as 
\beq
\mathcal{L} \supset -m_0^2 |X_i|^2 - B_0 W_2 - A_0 W_3\, .
\eeq
For an untwisted Polonyi field, characterized by the K\"ahler potential~(\ref{kahtuntwist1}), we find the following expressions for the induced soft terms
\begin{equation}
\begin{aligned}[c]
& {\rm Untwisted~Matter~Fields:} \\
& m_0^2 = \left(\alpha - 1 \right)m_{3/2}^2 , \\
& B_0 = -m_{3/2} , \\
& A_0 = 0,
\end{aligned}
\qquad \qquad
\begin{aligned}[c]
& {\rm Twisted~Matter~Fields:} \\
& m_0^2 = m_{3/2}^2 , \\
& B_0 =  -m_{3/2} , \\
& A_0 = 0 .
\label{softpoluntw}
\end{aligned}
\end{equation}
As one can see, the only dependence in the soft supersymmetry breaking terms on the curvature parameter $\alpha$
appears in the soft scalar masses for untwisted matter fields. 
When $\alpha = 1$, we have vanishing input scalar masses, which must then be generated by RGE evolution (typically above the GUT scale \cite{ENO5}).
When $\alpha = 2$, we obtain $m_0 = m_{3/2}$, $B_0 = -m_{3/2}$, and $A_0 = 0$, which is the pattern of soft terms when matter fields are twisted as well. In this case, we recover minimal supergravity (mSUGRA) \cite{bfs} boundary conditions, given by $A_0 = B_0 + m_0$, with $A_0 = 0$ as in models of pure gravity mediation (PGM) \cite{pgm,eioy}. 

In the untwisted case, imposing $\alpha \ge 1$ would avoid tachyonic soft supersymmetry-breaking scalar masses and the associated issue of vacuum stability. However, while this is condition is sufficient, it is not necessary~\cite{feng}. 
It is possible that soft supersymmetry-breaking scalar masses are negative at the input universality scale but no physical tachyonic scalars are found 
when the soft supersymmetry-breaking parameters are run
down to the weak scale. 
In fact, in studies of supersymmetric models with non-universal Higgs masses~\cite{nuhm2},
it was found that frequentist fits including many phenomenological and cosmological observables
were best fit with $m_0^2 < 0$ \cite{mc10}.
These models are however, 
potentially problematic due to the presence of 
charge- and/or colour-breaking minima~\cite{flat}.
However, if the electroweak vacuum is long-lived, the
relevance of other vacua becomes a cosmological
question related to our position in field space after inflation.
For a discussion of cosmological issues
associated with such tachyonic soft supersymmetry-breaking mass parameters,
see~\cite{EGLOS}.

For the case with a twisted Polonyi field, characterized by the K\"ahler potential~(\ref{kahttwist1}), we find 
\begin{equation}
\begin{aligned}[c]
& {\rm Untwisted~Matter~Fields:} \\
& m_0^2 = \alpha \, m_{3/2}^2 , \\
& B_0 = -m_{3/2} , \\
& A_0 = 0,
\end{aligned}
\qquad \qquad
\begin{aligned}[c]
& {\rm Twisted~Matter~Fields:} \\
& m_0^2 = m_{3/2}^2 , \\
& B_0 =  -m_{3/2} , \\
& A_0 = 0.
\end{aligned}
\end{equation}
The soft terms for twisted matter fields are unchanged from Eq. (\ref{softpoluntw})
and, once again, the only dependence on $\alpha$ appears for untwisted matter fields though, in this case, because $m_0 = \sqrt{\alpha} m_{3/2}$ 
there is no restriction on $\alpha$ other than its positivity. 

Note that we have not included here any modular weights in either the kinetic terms for twisted fields, or superpotential terms. These will be included in the next Section for $\phi$-type attractor models of inflation.  As explained in \cite{EGNO4}, the soft terms induced in $T$-type models are independent of all of the modular weights, which is not be the case for the $\phi$-type models, as we discuss in the next Section.~\footnote{However, the shifted minimum in Eq. (\ref{shiftone})
does depend on possible weights for the Polonyi field $Z$,
as discussed in Appendix~B.}

These key results are summarized in Table~\ref{ttype} below, and we briefly discuss cosmological aspects of such $T$-type models in Section~5. They were covered in detail in \cite{strongmoduli}.  

\begin{table}[ht!]
\centering
\renewcommand{\arraystretch}{1.5}
\setlength{\arrayrulewidth}{0.8pt}
\scalebox{1}{
\begin{tabular}{|
>{\columncolor[HTML]{C0C0C0}}c |c|c|c|c|}
\hline
\multicolumn{1}{|l|}{\cellcolor[HTML]{FFCCC9}{\color[HTML]{C0C0C0} }} & \multicolumn{2}{c|}{\cellcolor[HTML]{FFCCC9}\textit{\textbf{Untwisted Polonyi Field}}}                                  & \multicolumn{2}{c|}{\cellcolor[HTML]{FFCCC9}\textit{\textbf{Twisted Polonyi Field}}}                                     \\ \hline
\cellcolor[HTML]{C0C0C0}\textit{\textbf{VEVs}}                        & \multicolumn{2}{l|}{\cellcolor[HTML]{C0C0C0}}                                                                           & \multicolumn{2}{l|}{\cellcolor[HTML]{C0C0C0}}                                                                          \\ \hline
\cellcolor[HTML]{CBCEFB} \textit{\textbf{$\bm{\langle T \rangle}$} }                            & \multicolumn{2}{c|}{$\frac{1}{2} - \left(\frac{2 \alpha - 1}{ \alpha c^2}\right) \Delta^2$}                                     & \multicolumn{2}{c|}{$\frac{1}{2} + \frac{2 \alpha}{c^2} \Delta^2$}                                                       \\ \hline
\cellcolor[HTML]{CBCEFB} \textit{\textbf{$\bm{\langle \phi \rangle}$ }}                         & \multicolumn{2}{c|}{$\frac{\sqrt{3}}{c} \Delta$}                                                                 & \multicolumn{2}{c|}{$\frac{\sqrt{3 \alpha}}{c} \Delta$}                                                                  \\ \hline
\cellcolor[HTML]{CBCEFB} \textit{\textbf{$\bm{\langle Z \rangle}$}}                            & \multicolumn{2}{c|}{$\frac{\sqrt{\alpha}}{6 \sqrt{3}} \Lambda_Z^2$}                                                     & \multicolumn{2}{c|}{$\frac{1}{2 \sqrt{3}} \Lambda_Z^2$}                                                            \\ \hline
\cellcolor[HTML]{CBCEFB} \textit{\textbf{$\bm{b}$}}                                            & \multicolumn{2}{c|}{$\frac{1}{\sqrt{3 \alpha}} - \left( \frac{1 + 3 \alpha(\alpha - 1)}{2 \sqrt{3} \alpha^{3/2} c^2} \right) \Delta^2$}      & \multicolumn{2}{c|}{$\frac{1}{\sqrt{3}} - \frac{\sqrt{3} \alpha^2}{2 c^2} \Delta^2$}                                     \\ \hline
\textit{\textbf{$F$-term}}                    & \multicolumn{2}{c|}{$\frac{\mu}{\sqrt{\alpha}}$}                                                                                              & \multicolumn{2}{c|}{$\mu$}                                                                                \\ \hline
\textit{\textbf{$\bm{m_{3/2}}$}}              & \multicolumn{2}{c|}{$\frac{\mu}{\sqrt{3 \alpha}}$}                                                                             & \multicolumn{2}{c|}{$\frac{\mu}{\sqrt{3}}$}                                                                              \\ \hline
\textit{\textbf{$\bm{m_{z}}$}}                & \multicolumn{2}{c|}{$\frac{6 m_{3/2} }{\Lambda_Z}$}                                                    & \multicolumn{2}{c|}{$\frac{2 \sqrt{3} m_{3/2} }{\Lambda_Z}$}                                            \\ \hline
\cellcolor[HTML]{ACFF9E}\textit{\textbf{Matter Fields}}               & \cellcolor[HTML]{ACFF9E}\textit{\textbf{Untwisted}}      & \cellcolor[HTML]{ACFF9E}\textit{\textbf{Twisted}}            & \cellcolor[HTML]{ACFF9E}\textit{\textbf{Untwisted}}      & \cellcolor[HTML]{ACFF9E}\textit{\textbf{Twisted}}             \\ \hline

\textit{\textbf{$\bm{m_0^2}$}}                                        & $(\alpha - 1) m_{3/2}^2$                         &$~~~ m_{3/2}^2$                               & $\alpha \, m_{3/2}^2$                         & $ ~~~m_{3/2}^2$                                 \\ \hline
\textit{\textbf{$\bm{B_0}$}}                                          & $-m_{3/2}$ & $-m_{3/2}$ & $-m_{3/2}$ & $ -m_{3/2}$ \\ \hline
\textit{\textbf{$\bm{A_0}$}}                                          & $0$     & $0$     & $0$     & $0$      \\ \hline
\end{tabular}}
\caption{\it Model parameters and soft supersymmetry-breaking quantities in $T$-type no-scale attractor scenarios with either an untwisted or twisted Polonyi field and either untwisted or twisted
matter fields.}
\label{ttype}
\end{table}

\section{$\phi$-type No-Scale Attractor Scenarios}
\subsection{General Framework}
In this Section we discuss models of inflation where a matter-like field $\phi$ is interpreted as the inflaton.  
As discussed in~\cite{Avatars,rs,lrs}, inflationary models based on the minimal single field no-scale K\"ahler potential~(\ref{kah2}) entail uplifting a Minkowski vacuum via supersymmetry breaking,
which leads to an extremely heavy gravitino. An alternative way to construct viable inflationary models is to consider higher-dimensional non-compact coset manifolds~\cite{EKN2}, as mentioned in the Introduction.
There is a long history of constructing inflationary models this way \cite{GL,KQ,EENOS,otherns}. However, in many of the early models, the predictions of cosmological observables fall outside the range now determined by CMB observations \cite{planck18}. In \cite{ENO6},
it was shown that a simple Wess-Zumino superpotential can produce Starobinsky-like inflation, which leads to a spectral tilt, $n_s$, in good agreement with CMB measurements, and a tensor-to-scalar ratio, $r$, within reach of future experiments. The connection between Starobinsky inflation, $R + R^2$ gravity, and no-scale supergravity was further developed in \cite{eno9}. 

Here, we consider two possible non-minimal no-scale models, in which we introduce a single additional matter-like field, to be interpreted as the inflaton field. The inflaton may be included as an untwisted matter-like field, which 
parameterizes together with $T$ a non-compact $\frac{SU(2,1)}{SU(2) \cross U(1)}$ coset space~\cite{ENO6, Avatars}:
\begin{equation}
\label{kahuntwist}
{\rm Untwisted~Inflaton~Field:} \; \; K \; = \; -3 \, \alpha \ln \left( T + \overline{T} + \frac{\left(T+ \overline{T} - 1\right)^4}{\Lambda_T^2} + \frac{d\left(T - \overline{T} \right)^4}{\Lambda_T^2} - \frac{|\phi|^2}{3} \right),
\end{equation}
or as a twisted matter-like field, which parameterizes together with $T$ an $\frac{SU(1,1)}{U(1)} \cross U(1)$ space~\cite{EGNO123,ENNOV}:
\begin{equation}
\label{kahtwist}
{\rm Twisted~Inflaton~Field:} \; \; K \; = \; -3 \, \alpha \ln \left( T + \overline{T} + \frac{\left(T+ \overline{T} - 1\right)^4}{\Lambda_T^2} + \frac{d \left(T - \overline{T} \right)^4}{\Lambda_T^2}\right) + |\varphi|^2.
\end{equation}
In both cases, we include in the K\"ahler potential quartic stabilization terms for the volume modulus $T$, with $\Lambda_T < 1$. These stabilize the volume modulus $T$ in both the real and imaginary directions, and ensure that the VEV of the volume modulus is fixed dynamically to $\langle T \rangle = \frac{1}{2}$.

A superpotential that is a function only
of a matter-like field can be used to break supersymmetry and introduce a massive gravitino without invoking a hidden Polonyi sector \cite{EGNO4}. 
In such unified no-scale attractor models \cite{ENOV2,ENOV3}, the volume modulus $T$ plays the role of a Polonyi-like field that breaks supersymmetry, and only two complex fields are necessary for $\phi$-type models.~\footnote{In \cite{king1}, a term linear in $\phi$ is included, which plays the role of the Polonyi field, and Starobinsky-like inflation is possible so long as the gravitino mass $m_{3/2} \lesssim 1$ PeV.
The soft supersymmetry breaking parameters for this model were derived in \cite{king2}.} Furthermore, in addition
to inflation and supersymmetry breaking, these models can account for a small residual (though fine-tuned) cosmological constant.
The superpotential for such models can be written as~\cite{ENOV2}:
\begin{equation}
W \; = \; W_I + W_{dS},
\end{equation}
where $W_I$ characterizes inflation and $W_{dS}$ is responsible for supersymmetry breaking and a (small) positive cosmological constant that appears at the end of inflation \cite{EKN2,ENNO}. The forms of the superpotentials
$W_I$ and $W_{dS}$ depend whether the inflaton is twisted or untwisted and therefore combined with the K\"ahler potential in either Eq.~(\ref{kahuntwist}) or~(\ref{kahtwist}) respectively~\cite{ENNOV}:
\begin{align}
{\rm Untwisted~Inflaton~Field:} \; \; &
\label{untsup1}
W_{I} = \sqrt{\alpha} \, M \, f(\phi)  \cdot \left(2T - \frac{\phi^2}{3} \right)^{n_{-}}, \\
 &
 \label{untsup2}
 W_{dS}  =  \lambda_1 \, M^3 \cdot \left(2T - \frac{\phi^2}{3} \right)^{n_{-}} - \, \lambda_2 \, M^3 \cdot \left(2T - \frac{\phi^2}{3} \right)^{n_{+}},
 \end{align}
 and
 \begin{align}
{\rm Twisted~Inflaton~Field:} \; \; & \label{tsup1} 
W_{I}  = M  \, f(\varphi) \cdot \left(2T \right)^{n_{-}} \cdot e^{-\frac{\varphi^2}{2}} , \\ 
& \label{tsup2} 
W_{dS} =  \left[\, \lambda_1 \, M^3 \cdot \left(2T \right)^{n_{-}} - \, \lambda_2 \, M^3 \cdot \left(2T\right)^{n_{+}} \right] \cdot e^{-\frac{\varphi^2}{2}},
\end{align}
where $n_{\pm} = \frac{3}{2} \left(\alpha \pm \sqrt{\alpha} \right)$,~$M \simeq 1.2 \times 10^{-5} M_P \simeq 3 \times 10^{13} \, \rm GeV$ is the inflaton mass for Starobinsky-like inflation, and one of the couplings $\lambda_{i}$ must be tuned to a obtain a small vacuum density, whereas the other may be of order 1. 

After the volume modulus $T$ is stabilized by the quartic terms in the K\"ahler potential forms~(\ref{kahuntwist}) and~(\ref{kahtwist}), with a VEV~$\langle T \rangle = \frac{1}{2}$,  the inflaton field $\phi$ is stabilized in the imaginary direction throughout inflation in both cases, and we have $\phi = \bar{\phi}$. Note that, 
despite the presence of supersymmetry breaking and a non-zero final vacuum energy density, Starobinsky-inflation is reproduced for an appropriate choice of $f(\phi)$. 

If we combine the K\"ahler potential~(\ref{kahuntwist}) with the superpotentials~(\ref{untsup1}) and~(\ref{untsup2}), the effective scalar potential~(\ref{effpot}) becomes
\begin{equation}
\label{pot1}
{\rm Untwisted~Inflaton~Field:} \; \; V \simeq \Lambda + M^2 \left(1 - \frac{\phi^2}{3} \right)^{1 - 3 \sqrt{\alpha}} f'(\phi)^2,
\end{equation}
and similarly, if we combine the K\"ahler potential~(\ref{kahtwist}) with
the superpotentials~(\ref{tsup1}) and~(\ref{tsup2}), equation~(\ref{effpot}) gives
\begin{equation}
\label{pot2}
{\rm Twisted~Inflaton~Field:} \; \; V \simeq \Lambda +M^2 \, f'(\varphi)^2,
\end{equation}
where
\begin{equation}
\label{cc1}
\Lambda = 12 \, \lambda_1 \, \lambda_2 \, M^6.
\end{equation}
In both cases, the cosmological constant $\Lambda$ depends on two constants, $\lambda_{1,2}$. It should be noted that we neglected the contribution of a term which is proportional to  $\lambda_{2} M^4 f $. This term vanishes at the minimum and, because
$\lambda_2 M^4 \ll M^2$, it does not affect the inflationary dynamics \cite{ENOV3}.

Being proportional to $M^6$, the cosmological constant is of order $10^{-30} \lambda_2$ in Planck units, when we assume $\lambda_1 \sim \mathcal{O}(1)$.
We expect that the final vacuum energy density is modified by (negative) contributions from
phase transitions occurring after inflation.
For $\lambda_2\sim \mathcal{O}(1)$, we would require a contribution of order $M^6 \sim 10^{-30}$ to cancel the term in (\ref{cc1})
to eventually yield a cosmological constant of order $10^{-120}$ today. For example, 
the GUT phase transition in a flipped SU(5) $\times$ U(1) model occurs after inflation \cite{EGNNO23,EGNNO45} and contributes
$\Delta V \sim - M_{\rm susy}^2 M_{\rm GUT}^2 \sim - ({\lambda}_1 - {\lambda}_2)^2 M^6 M_{\rm GUT}^2$, indicating that perhaps ${\lambda_2} \sim  (M_{\rm GUT}/M_P)^2$ for $\lambda_1 \sim 1$.

In the untwisted case, the $\alpha$-Starobinsky inflationary potential 
\begin{equation}
V \simeq \Lambda + \frac{3}{4} M^2 \left(1 - e^{- \sqrt{\frac{2}{3 \alpha}} x} \right)^2,
\label{astaroL}
\end{equation}
can be obtained from Eq.~(\ref{pot1}) with the choice of $f(\phi)$ which satisfies \cite{ENOV3}
\beq
 f'(\phi ) = \frac{\sqrt{3} \,\phi}{\left(\phi +\sqrt{3}\right)}   \left(1-\frac{\phi ^2}{3}\right)^{(3 \sqrt{\alpha }-1)/2}  \, ,
 \label{f'}
\eeq
and a field redefinition
\begin{equation}
\phi = \sqrt{3} \tanh(\frac{x}{\sqrt{6 \alpha}}).
\label{kinphi}
\end{equation}
The superpotential function $f(\phi)$ derived from Eq.~(\ref{f'}) with boundary condition $f(0) = 0$ is in general a hypergeometric function, which assumes a polynomial form whenever
$9 \alpha$ is an odd perfect square other than 1.
For example, when $\alpha = 1$, $f(\phi)$ is of the Wess-Zumino form \cite{ENO6}
\begin{equation}
f(\phi) = \left(\frac{\phi^2}{2} - \frac{\phi^3}{3 \sqrt{3}} \right) \, .
\label{wi}
\end{equation}
In the twisted case, the $\alpha$-Starobinsky inflationary potential (\ref{astaroL})
can be obtained from Eq.~(\ref{pot2}) with the choice of $f(\varphi)$ which satisfies 
\beq
 f'(\varphi ) = \frac{\sqrt{3}}{2} \left(1 - e^{-\frac{2 \varphi}{\sqrt{3 \alpha}}} \right) \, ,
 \label{f't}
\eeq
and a field redefinition
\begin{equation}
\varphi = x/\sqrt{2}.
\label{kinvarphi}
\end{equation}
In this case, there is a relatively simple form for $f(\varphi)$ for all $\alpha$
\begin{equation}
f(\varphi) =  \frac{3\sqrt{\alpha}}{4} \left(\frac{2 \varphi }{\sqrt{3\alpha}}+e^{-\frac{2 \varphi }{\sqrt{3}}}-1\right) \, .
\label{wvari}
\end{equation}

\subsection{Supersymmetry Breaking}

The unified no-scale attractor models with an untwisted or a  twisted inflaton field both
yield a de Sitter vacuum at the minimum, and the two formulations can be considered as equivalent
for cosmological purposes. At the end of inflation, supersymmetry is broken through an $F$-term for $T$, which is given by~\cite{ENOV2, ENOV3}
\begin{equation}
\label{fterm}
\sum_{i = 1}^{2} |F_i|^2 = F_T^2 \simeq \frac{ \left(\lambda_1 + \lambda_2 \right)^2}{\alpha} M^6,
\end{equation}
and the gravitino mass is given simply  by
\begin{equation}
\label{grav}
m_{3/2} = e^{G/2} = e^{K/2} W = \left(\lambda_1 - \lambda_2 \right) \frac{M^3}{M_P^2},
\end{equation}
which is independent of the curvature parameter $\alpha$. Thus, in our framework 
the $F$-term is a function of the curvature $\alpha$ whereas the gravitino mass $m_{3/2}$ is not. 
Moreover, as we have mentioned before, our framework incorporates supersymmetry breaking without introducing an additional Polonyi sector~\cite{pol} or external uplifting by fibres~\cite{KKLT}. 

In order to obtain a gravitino mass $m_{3/2} \simeq \mathcal{O} (1) \, \text{TeV}$, we choose $\lambda_2 \ll \lambda_1$. Then we can write
\begin{equation}
\label{lambda}
m_{3/2} = \left(\lambda_1 - \lambda_2 \right) \frac{M^3}{M_P^2} \simeq \lambda_1 \frac{M^3}{M_P^2} \, ,
\end{equation}
and we can re-express the $F$-term for $T$~(\ref{fterm}) as $F_T \simeq \frac{m_{3/2} }{\sqrt{\alpha}}$.
We note that, by scaling $W_{dS}$ with $M^3$,
we obtain a TeV mass scale for supersymmetry breaking 
without fine-tuning, and relate the supersymmetry-breaking
scale to the inflation scale $M$ (see also \cite{dgmo}).

Using (\ref{grav}) and~(\ref{lambda}), we find that the squared masses of the real and imaginary components of the volume modulus $T$ are given by:
\begin{equation}
\label{massT}
m_{\rm{Re \, T}}^2 \simeq  \frac{48 \, m_{3/2}^2}{\alpha \Lambda_T^2}, \qquad m_{\rm{Im \, T}}^2 \simeq  \frac{48 d m_{3/2}^2}{\alpha \Lambda_T^2},
\end{equation}
which depend on the supersymmetry-breaking parameter $\lambda_1$, the stabilization constant $\Lambda_T$ and the curvature parameter $\alpha$. For $\Lambda_T \ll 1$, we have a built-in hierarchy between the modulus and the gravitino mass scales.  It is important to note that, in the absence of supersymmetry breaking, $m_{3/2} = 0$, and both $T$ components remain massless.

Finally, the squared mass of the inflaton is given by:
\begin{equation}
\label{massphi}
m_{\phi}^2 \simeq M^2 \, f''(0)^2,
\end{equation}
where we may assume that $f''(0) \sim \mathcal{O}(1)$, in which case the inflaton mass is $m_{\phi} = M \simeq \mathcal{O} (10^{-5})$. 

\subsection{Incorporating Matter Particles}

In order to incorporate Standard Model-like particles in $\phi$-type models of unified no-scale attractors, we illustrate different possible superpotential structures that couple the hidden and visible matter sectors. 

We consider the following general superpotential form
\begin{equation}
W = W_I + W_{dS} + W_{SM},
\end{equation}
where in the case of an untwisted inflaton, the superpotentials $W_I$ and $W_{dS}$ are given by~(\ref{untsup1}) and~(\ref{untsup2}) and  for a twisted inflaton field $W_I$ and $W_{dS}$ are given by~(\ref{tsup1}) and~(\ref{tsup2}), and $W_{SM}$ describes the Standard Model-like interactions, given by:
\begin{equation}
\label{sm1}
W_{SM}  = \sqrt{\alpha} \left[\left(T + \frac{1}{2} \right)^{\beta} W_2 (X_i) + \left(T + \frac{1}{2} \right)^{\gamma} W_3 (X_i)\right]
 \cdot Y^{n_{-}},
\end{equation}
where we have introduced bilinear and trilinear couplings $W_{2,3}$ with non-zero modular weights $\beta$ and $\gamma$, and
\begin{align}
{\rm Untwisted~Inflaton~Field:} & \qquad Y = 2T - \frac{\phi^2}{3}, \\
{\rm Twisted~Inflaton~Field:} & \qquad Y = 2T.
\end{align}
It should also be noted that we couple the Standard Model-like sector to $(2T - \phi^2/3)^{n_{-}}$ for the case with an untwisted inflaton field and $(2T)^{n_{-}}$ for the case with a twisted inflaton field, where $n_{-} = \frac{3}{2} \left(\alpha - \sqrt{\alpha} \right)$. One may also consider couplings to $Y^{n_{+}}$, where $n_{+} = \frac{3}{2} \left(\alpha + \sqrt{\alpha} \right)$, and find similar results.

As in the previous Section, matter fields may appear either as untwisted in the K\"ahler potential as in Eq.~(\ref{untwisted}) or 
as twisted fields in the K\"ahler potential
\beq
K \supset |X_i|^2 (T +\overline{T})^{-n_i} \, ,
\label{ni}
\eeq
which sits outside the logarithmic term and
where we have included a kinetic modular weight, $n_i$.
For $T$-type inflation, soft mass terms do not depend on modular weights and so these were neglected in writing the superpotential in Eq. (\ref{WType}). On the other hand, in $\phi$-type models the modular weights do enter 
into the soft supersymmetry breaking terms, and the
weights $\beta$ and $\gamma$ are included in Eq. (\ref{sm1}) separately for bilinear and trilinear couplings. 
We obtain the following induced soft terms for untwisted matter fields~(\ref{untwisted}) and twisted matter fields~(\ref{twisted}):
\begin{equation}
\label{termsphi}
\begin{aligned}[c]
& {\rm Untwisted~Matter~Fields:} \\
& m_0^2 = \left(\alpha - 1 \right) m_{3/2}^2 , \\
& B_0 = \left(2 \sqrt{\alpha} - 2 - \beta \right) m_{3/2} , \\
& A_0 = \left(3 \sqrt{\alpha} - 3 - \gamma \right) m_{3/2} ,
\end{aligned}
\qquad \qquad
\begin{aligned}[c]
& {\rm Twisted~Matter~Fields:} \\
& m_0^2 = \frac{(\alpha - n_i)}{\alpha} m_{3/2}^2 , \\
& B_0 = \left(2 \sqrt{\alpha}  - 2 n_i - \beta \right) m_{3/2} , \\
& A_0 = \left(3 \sqrt{\alpha} - 3 n_i - \gamma \right) m_{3/2} .
\end{aligned}
\end{equation}
For $\alpha = 1$, these results reduce to those found in \cite{EGNO4}. 

The induced soft terms~(\ref{termsphi}) allow us to consider various phenomenological scenarios. Let us first consider $\alpha = 1$. For untwisted matter fields,  we obtain $m_0 = 0$, $B_0 = -\beta m_{3/2}$, and $A_0 = -\gamma m_{3/2}$. If we set $\beta = \gamma = 0$, we recover standard no-scale soft terms with $A_0 = B_0 = m_0 = 0$~\cite{no-scale,Ellis:1983sf,EKN1}.  
For twisted matter fields with $n_i = 0$ or universal,
one finds non-zero universal soft mass terms as in the 
 constrained minimal supersymmetric extension of the Standard Model (CMSSM) \cite{cmssm}, and if $\beta = \gamma$ one obtains soft terms of the mSUGRA type with $A_0 = B_0 + m_0$, all proportional to the gravitino mass.
 If $\beta$ and $\gamma$ vanish, we have $A_0 = 3 m_0$
 and $B_0 = 2 m_0$. For $\beta  = \gamma = 3$,
 we obtain PGM~\cite{pgm,eioy}
 soft terms, given by $m_0 = m_{3/2}$,  $A_0 = 0$ and $B_0 = -m_{3/2}$. Finally, we note
 that scalar mass universality is lost if the modular weights $n_i$ are not universal. 
 
As in the case of $T$-type models, for untwisted matter fields with $\alpha < 1$ or for twisted matter fields with $\alpha < n_i$, there is the possibility that $m_0^2 < 0$. However, as discussed above, such models are not necessarily excluded by cosmological considerations. For $\alpha > 1$, one finds non-zero scalar masses even in the untwisted case. 
 Finally, we point out that these results do not depend whether the inflaton is twisted or not. 
Our key results for $\phi$-type unified no-scale models of inflation are summarized in Table~\ref{noscaleTable}.

\begin{table}[h!]
\centering
\renewcommand{\arraystretch}{1.5}
\setlength{\arrayrulewidth}{0.8pt}
\begin{tabular}{|
>{\columncolor[HTML]{C0C0C0}}c |c|c|}
\hline    
\cellcolor[HTML]{FFCCC9}                          & \multicolumn{2}{c|}{\cellcolor[HTML]{FFCCC9}\textit{\textbf{Twisted/Untwisted Inflaton Field}} }                                                                  \\ \hline
\textit{\textbf{$F$-term}}                             & \multicolumn{2}{c|}{$\frac{m_{3/2}}{\sqrt{\alpha}}$}                                                                  \\ \hline
\textit{\textbf{$\bm{m_{\rm \, Re \, T}^2}$}}             & \multicolumn{2}{c|}{$\frac{48 m_{3/2}^2}{\alpha \Lambda_T^2}$}                                                        \\ \hline
\textit{\textbf{$\bm{m_{\rm \, Im \, T}^2}$}}             & \multicolumn{2}{c|}{$\frac{48 d m_{3/2}^2}{\alpha \Lambda_T^2}$}                                                      \\ \hline
\cellcolor[HTML]{ACFF9E}\textit{\textbf{Matter Fields}}                        & \cellcolor[HTML]{ACFF9E}\textit{\textbf{Untwisted}}        & \cellcolor[HTML]{ACFF9E}\textit{\textbf{Twisted}}        \\ \hline

\cellcolor[HTML]{C0C0C0}\textit{\textbf{$\bm{m_0^2}$}} & $\left( \alpha -1 \right) m_{3/2}^2$              & $\frac{(\alpha - n_i)}{\alpha} m_{3/2}^2$                \\ \hline
\cellcolor[HTML]{C0C0C0}\textit{\textbf{$\bm{B_0}$}}   & $\left(2 \sqrt{\alpha} - 2 - \beta \right) m_{3/2}$        & $\left(2 \sqrt{\alpha} - 2 n_i - \beta \right) m_{3/2}$  \\ \hline
\cellcolor[HTML]{C0C0C0}\textit{\textbf{$\bm{A_0}$}}   & $\left(3 \sqrt{\alpha} - 3 - \beta \right) m_{3/2}$ & $\left(3 \sqrt{\alpha} - 3 n_i - \gamma \right) m_{3/2}$ \\ \hline
\end{tabular}
\caption{\it Model parameters and soft supersymmetry-breaking quantities in $\phi$-type unified no-scale attractor scenarios with either untwisted or twisted
matter fields. The quantities $\beta, \gamma$ and $n_i$ are modular weights introduced in (\ref{sm1}) and (\ref{ni}), respectively.}
\label{noscaleTable}
\end{table}

\section{Cosmological Scenarios, Entropy and Dark Matter Production}

Supersymmetry breaking has often been a source of cosmological Angst. The so-called Polonyi problem, or more generally the moduli problem, arises when scalars with weak scale masses but with Planck scale vacuum expectation values are displaced from their minima after inflation \cite{polprob}. Their evolution and late decay generally
produce enormous amounts of entropy, washing away any baryon asymmetry. Their decays into supersymmetric particles may also lead to an excessive dark matter abundance \cite{myy,kmy} in the form of the lightest supersymmetric particle (LSP), if R-parity is preserved. 
In this Section we discuss these cosmological issues in $T$- and $\phi$-type no-scale attractor models. 

\subsection{Post-inflationary Dynamics in $T$-Type Models}
As we have seen, in $T$-type models inflation is driven by a volume modulus $T$ whose dynamics is characterized by a function $f(T)$, and supersymmetry is broken through a Polonyi field $Z$. When the inflaton rolls down to a Minkowski minimum, given by the left side of~(\ref{shiftone}) for the case with untwisted Polonyi field, and by the right side of~(\ref{shiftone}) for the case with a twisted Polonyi field, the fields (both the inflaton and $Z$) begin to oscillate and their subsequent decay begins the process of  reheating of the Universe. For example, during inflation, a twisted Polonyi field $Z$ is displaced to a minimum determined by Hubble induced mass corrections $\sim H_I^2 Z \bar{Z}$, which is much smaller than the VEV of $\langle Z \rangle$ at the true minimum, i.e., $\langle Z \rangle_{\rm Inf} \ll \langle Z \rangle$~\cite{moduli1, moduli2, strongmoduli}. When the Hubble parameter becomes smaller than the Polonyi mass, $m_z$, more precisely when $H \lesssim \frac{2}{3} m_z$, the inflationary minimum of $Z$ starts moving adiabatically toward the true minimum. As a result, the initial amplitude of the field is very small and roughly proportional to 
\beq
\langle z \rangle_{\rm Max} \sim \Lambda_Z^2 \, .
\label{polmax}
\eeq
The decay of the inflaton is model-dependent, but the main decay channel of the Polonyi field $Z$ is into a pair of gravitinos, potentially exacerbating the gravitino problem \cite{modgrav}. The strongest limit on $\Lambda_Z$ comes from the decay of Z to gravitinos and their subsequent decay to the LSP, $\chi$.  In \cite{strongmoduli}, it was found that 
$\Lambda_Z \lesssim 3 \times 10^{-4}$ for $m_\chi = 100$ GeV, 
and $m_{3/2} = 10^{-15} M_P$, and scales as $(m_{3/2}/m_\chi)^{1/5}$. A more complete and detailed treatment of cosmological consequences of $T$-type models is presented in~\cite{strongmoduli}. The results derived there apply to the $T$-type models discussed here, and we do not discuss them further in this paper.

\subsection{Post-inflationary Dynamics in $\phi$-Type Models}
As we have also seen, in $\phi$-type models inflation is driven by a matter-like field $\phi$, and inflationary dynamics is characterized by a function $f(\phi)$~\cite{ENOV2}. In this case, supersymmetry is broken by the volume modulus $T$, which acts like a Polonyi field. The inflaton field $\phi$ exits the high-lying de Sitter plateau and rolls down toward a low-lying global de Sitter minimum, characterized by $\Lambda  = 12 \, \lambda_1 \, \lambda_2 \, M^6$. When the inflaton reaches this de Sitter minimum, located at $\langle T \rangle = \frac{1}{2}$ and $\langle \phi \rangle = 0$, it starts oscillating about the minimum, with an initial maximum amplitude given by
\begin{equation}
\label{osc}
\langle T \rangle_{\rm{Max}} \simeq \frac{\Lambda_T}{4 \sqrt{3}}, \qquad \langle \phi \rangle_{\rm{Max}} \simeq \mathcal{O} (10^{-1}).
\end{equation}
Note the magnitude of the initial amplitude of $T$ oscillations is a key difference between these models and the $T$-type models. In the latter, the maximum amplitude for 
strongly-stabilized Polonyi oscillations, given in Eq.~(\ref{polmax}) is $\propto \Lambda_Z^2$. In this case, the initial amplitude is significantly larger, and we expect stronger constraints on $\Lambda_T$.
The main decay channel of the volume modulus $T$ is into a pair of gravitinos.

We consider now various post-inflationary scenarios in $\phi$-type models,
and calculate the corresponding upper limits on the modulus stabilization parameter $\Lambda_T$. At the end of the inflationary epoch, when the inflaton rolls down toward a global minimum and starts oscillating about it, and the Universe enters a period of matter-dominated expansion. As the Hubble damping parameter decreases, the volume modulus $T$ starts oscillating about its minimum. These oscillations may occur before or after reheating of the Universe, and we treat these cases separately below. 

When the inflationary period is over (i.e., the near-exponential expansion ends), the energy density of the Universe is dominated by the oscillations of the inflaton about its minimum.
We define the start of inflaton oscillations as the moment when the Hubble parameter is of order the inflaton mass, or more precisely when  $H = \frac23 m_\eta$, where  $\eta$ ($= \phi$ in the untwisted case or $\varphi$ in the twisted case) is the inflaton.
If we define $R_\eta$ as the scale factor when inflaton oscillations begin,
we can write the energy density and Hubble parameter as
\begin{equation}
\label{energy1}
\rho_{\eta} \simeq \frac{4}{3} m_{\eta}^2 M_P^2 \left( \frac{R_{\eta}}{R} \right)^3,
\end{equation}
\begin{equation}
\label{hubble1}
H \simeq \frac{2}{3} m_{\eta} \left(\frac{R_{\eta}}{R} \right)^{3/2}.
\end{equation}
During this matter-dominated epoch, the Hubble expansion rate keeps decreasing until the field $T$ begins oscillating about the minimum. These oscillations start when the Hubble parameter approaches the mass of $T$ or when $H \simeq \frac23 m_{T}$, where $m_T$ is the canonically-normalized mass of the volume modulus. This value of the Hubble parameter can be reached before or after the inflaton $\eta$ decays, i.e., before or after reheating. When the field $T$ begins oscillating about the supersymmetry-breaking minimum, the energy density of oscillations is given by
\begin{equation}
\label{energy2}
\rho_T \simeq \frac{1}{2} m_T^2 M_P^2 \langle T \rangle_{\rm{Max}}^2 \left( \frac{R_T}{R} \right)^3,
\end{equation}
where $\langle T \rangle_{\rm{Max}}$ is the maximum amplitude of the oscillations of the volume modulus and $R_T$ is the cosmological scale factor at the onset of $T$ oscillations, which may begin before or after inflaton decays. 

We distinguish the following scenarios for the possible evolution of the Universe.

\textbf{Scenario I}: {\it Oscillations of the volume modulus $T$ begin before inflaton decay}

We assume that the inflaton couples only through gravitational-strength interactions, in which case the inflaton decay rate can be estimated as
\begin{equation}
\label{decayphi}
\Gamma_{\eta} = {d_{\eta}^2 \frac{m_{\eta}^3}{M_P^2}},
\end{equation}
where $d_{\eta}$ is a gravitational-strength coupling \footnote{If there are non-gravitational couplings leading to inflaton decay, we can write $d_\eta = {\tilde d}_\eta M_P/m_\eta$, where ${\tilde d}_\eta$ is the non-gravitational coupling.}. The inflaton decays when the Hubble parameter decreases to the critical value $\Gamma_{\eta} = \frac{3}{2} H$ where, using the expressions~(\ref{hubble1}) and~(\ref{decayphi}), we obtain the following scale factor ratio
\begin{equation}
\label{scale2}
\frac{R_{d \eta}}{R_{\eta}} = \left(\frac{M_P}{d_{\eta} m_{\eta}} \right)^{4/3},
\end{equation}
where $R_{d \eta}$ is the cosmological scale factor at the time of inflaton decay.

Using equations~(\ref{massT}, \ref{osc}, \ref{energy2}), we can express the energy density stored in $T$ oscillations~(\ref{energy2}) as
\begin{equation}
\label{energy22}
    \rho_T \simeq \frac{1}{2 \alpha} m_{3/2}^2 M_P^2 \left( \frac{R_T}{R} \right)^3.
\end{equation}
To find the scale factor at the beginning of $T$ oscillations, we use~(\ref{massT}, \ref{hubble1}) and $H \simeq \frac{2}{3} m_T$ to estimate~\footnote{It should be noted that $T$ oscillations might also begin after the inflaton decays, and we discuss this possibility later.}
\begin{equation}
\label{scaleT}
R_T \simeq  \alpha^{1/3} \left(\frac{\Lambda_T}{4 \sqrt{3}} \frac{m_{\eta}}{m_{3/2}} \right)^{2/3} R_{\eta}.
\end{equation}
We seek the limits on $\Lambda_T$ that ensure that entropy production from $T$ decay does not cause excessive dilution of the primordial baryon-to-entropy ratio, $n_B/s \simeq 8.7 \times 10^{-11}$.

In this scenario, we are assuming that oscillations of the volume modulus $T$ begin before inflaton decay, i.e., $R_{T} < R_{d \eta}$. Using the scale factor ratios~(\ref{scaleT}) and~(\ref{scale2}), we find
\begin{equation}
\frac{R_T}{R_{d \eta}} = \alpha^{1/3} \left(\frac{d_{\eta}^2 \Lambda_T M_P}{4 \sqrt{3} m_{3/2}} \right)^{2/3} \left(\frac{m_{\eta}}{M_P}\right)^2 < 1,
\end{equation}
which leads to the following condition for Scenario I 
\begin{equation}
\label{lambda1}
    \Lambda_T \lesssim \alpha^{-1/2} \frac{4 \sqrt{3} m_{3/2}}{d_{\eta}^2} \frac{M_P^2}{m_{\eta}^3}.
\end{equation}
Therefore, if we choose~$m_{\eta} \sim 10^{-5} M_P$, $m_{3/2} \sim 10^{-15} M_P \sim \mathcal{O}(1) \, \rm{TeV}$, and $\alpha = 1$, we find $T$ oscillations begin before inflaton decay when $d_{\eta}^2 \Lambda_T \lesssim \sqrt{48}$, a condition which is almost always satisfied. We must now distinguish between the possibilities that $T$ decays before the inflaton (I~a), after the inflaton but before $T$ oscillations dominate the energy density (I~b), and when they do dominate the energy density (I~c). 

We assume instantaneous inflaton decay and thermalization of inflaton decay products. 
When the inflaton decays the Universe becomes radiation-dominated, and the energy density and Hubble parameter have the following expressions
\begin{equation}
\label{energy3}
\rho_r = \frac{4}{3} d_{\eta}^{-4/3} m_{\eta}^{2/3} M_P^{10/3} \left(\frac{R_{\eta}}{R} \right)^4,
\end{equation}
\begin{equation}
\label{hubble3}
H = \frac{2}{3} d_{\eta}^{-2/3} m_{\eta}^{1/3} M_P^{2/3} \left(\frac{R_{\eta}}{R} \right)^2 \, ,
\end{equation}
and the reheating temperature is given by
\begin{equation}
\label{reheat}
T_{\rm{RH}} = d_{\eta} \left( \frac{40}{\pi^2 g_{\eta}} \right)^{1/4} \frac{m_{\eta}^{3/2}}{M_P^{1/2}},
\end{equation}
where $g_{\eta} = g(T_{\rm RH})$ is the effective number of degrees of freedom at reheating, for which we take the MSSM value $g(T_{\rm{RH}}) = 915/4$. We note that, in this model, the reheating temperature is model-dependent and depends on how the inflaton 
is coupled to the Standard Model, and hence also its decay rate defined in
Eq.~(\ref{decayphi}). The preferred values of $N_*$ and $n_s$ also depend on the decay rate, though very weakly as discussed in detail in \cite{reheating}.

To calculate the decay rate of the volume modulus to gravitinos, we use the following fermion-scalar interaction supergravity Lagrangian (see e.g. \cite{Nilles:1983ge}), where we work in the unitary gauge:
\begin{equation}
\label{lagsup}
    \mathcal{L}_{\rm{F, int}} \supset \frac{i}{2} \bar{\chi}_{\rm{i L}} \slashed{D} \Phi_j \chi_{L}^{k} \left(-G^{i j}_k + \frac{1}{2} G^i_k G^j \right) + \frac{1}{2}e^{G/2} \left(-G^{ij} - G^i G^j + G^{i j}_k (G^{-1})^k_l G^l \right) \bar{\chi}_{\rm{iL}} \chi_{\rm{jR}} + \rm{h.c.} \, .
\end{equation}
In this case, the modulino $\chi_T$ becomes the longitudinal component of the gravitino, and the dominant two body decay $T \rightarrow \psi_{3/2} \psi_{3/2}$ coupling can be obtained from the following interaction term in the Lagrangian
\begin{equation}
\mathcal{L}_{F, 2} \supset \frac{1}{2} m_{3/2} G_{T}^{TT} (G^{-1})^T_T G^T \bar{\psi}_{3/2} \psi_{3/2} = 36 \alpha \frac{m_{3/2}}{M_P \Lambda_T^2} \overline{T} \bar{\psi}_{3/2} \psi_{3/2}\, .
\end{equation}
The decay rate obtained from this interaction term is given by
\begin{equation}
\label{decayT1}
\Gamma_T^{\rm{(Total)}} \simeq \Gamma(T \rightarrow \psi_{3/2} \psi_{3/2}) = \alpha^{3/2} \frac{648 \sqrt{3} m_{3/2}^3}{\pi M_P^2 \Lambda_T^5} \, .
\end{equation}
We now discuss in turn the specific cases I~a), b), c) mentioned above.

\textbf{I a):} {\it The volume modulus $T$ decays before the inflaton} 

In this scenario, the decay of the field $T$ is characterized by $\Gamma_T = \frac{3}{2} H$, and the Hubble parameter for the matter-dominated Universe is given by~(\ref{hubble1}):
\begin{equation}
\label{scale5}
\frac{R_{dT}}{R_{\eta}} = \frac{1}{108 \alpha} \left(\frac{\pi  m_{\eta} \Lambda_T^{5} M_P^2}{m_{3/2}^3} \right)^{2/3},
\end{equation}
where $R_{dT}$ is the cosmological scale factor when the volume modulus $T$ decays. 
To obtain the case when the volume modulus $T$ decays before $\eta$, we must impose $R_{dT} < R_{d \eta}$, and if we use the scale factor ratio~(\ref{scale2}) with~(\ref{scale5}), we find the following upper bound for $\Lambda_T$:
\begin{equation}
\Lambda_T \lesssim \, 3.2 \, \alpha^{3/10} \left(\frac{m_{3/2}^3}{d_\eta^2 m_{\eta}^3} \right)^{1/5} \, .
\label{IaLT}
\end{equation}
The representative values $m_{\eta} \sim 10^{-5} M_P$, $m_{3/2} \sim 10^{-15} M_P$, $d_\eta \sim 1$, and $\alpha = 1$ yield $\Lambda_T < 3.2 \times 10^{-6}$.
In this case the entropy released after $T$ decay is negligible.

\textbf{I b):} {\it The volume modulus $T$ decays after the inflaton, but never dominates}

If the decay of $T$ occurs after inflaton reheating, the value of the scale factor at 
$T$ decay is computed using the radiation-dominated form of $H$ given in Eq.~(\ref{hubble3}). Then, using $\Gamma_T = 2 H_r$,  we find following scale factor ratio:
\begin{equation}
\frac{R_{dT}}{R_{\eta}} = \alpha^{-3/4} \left(\frac{\pi}{486 \sqrt{3}}\right)^{1/2} \frac{   m_{\eta}^{1/6} \Lambda_T^{5/2} M_P^{4/3}}{ d_{\eta}^{1/3} m_{3/2}^{3/2} }\, .
\label{Rdtder}
\end{equation}
In order to realize this scenario,
we must impose $R_{dT} > R_{d\eta}$,
which holds when Eq.~(\ref{IaLT}) is violated. 
However in this case, we are also assuming that $T$ never comes to dominate the energy density, i.e.,
$\rho_T(R_{dT}) < \rho_r (R_{dT})$.
From the expressions above, we have
\beq
\frac{\rho_T}{\rho_r} = \alpha^{-3/4} \frac{d_\eta}{128} \left(\frac{\pi}{486\sqrt{3}}\right)^{1/2} 
\left(\frac{m_\eta \Lambda_T^3}{ m_{3/2}}\right)^{3/2}< 1
\eeq
which gives an upper bound on $\Lambda_T$:
\beq
\Lambda_T < 5.5 \alpha^{1/6} d_\eta^{-2/9}  \left(\frac{m_{3/2}}{m_\eta}\right)^{1/3} ,
\label{IbLT}
\eeq
so that for $m_{\eta} \sim 10^{-5} M_P$, $m_{3/2} \sim 10^{-15} M_P$, $d_\eta \sim 1$, and $\alpha = 1$, $\Lambda_T < 2.5 \times 10^{-3}$.

\textbf{I c):} {\it The volume modulus $T$ decays after the inflaton, and dominates at decay}

For $\Lambda_T$ larger than the upper limit in Eq.~(\ref{IbLT}),  oscillations of the volume modulus $T$ will dominate the energy density, i.e., $\rho_T > \rho_{r}$ before decay. 
In this case, the Hubble parameter is given by
\begin{equation}
\label{hubbleT}
    H_T = \frac{1}{\sqrt{6 \alpha}} m_{3/2} \left(\frac{R_T}{R} \right)^{3/2},
\end{equation}
as the Universe becomes matter-dominated again. 

The volume modulus $T$ decays when $H_T \simeq \frac{2}{3} \Gamma_T$, and using the expressions~(\ref{hubbleT}) and~(\ref{decayT1}), we find the following scale factor ratio
\begin{equation}
\label{scale3}
\frac{R_{dT}}{R_{T}} = \alpha^{-4/3} \left(\frac{\pi}{1296 \sqrt{2} } \frac{\Lambda_T^5 M_P^2}{m_{3/2}^2} \right)^{2/3}.
\end{equation}
Because $T$ dominates when it decays, its decay products  increase the entropy. 
The entropy densities in radiation and $T$ are given by
\begin{equation}
\label{entropy}
s_r = \frac{4}{3} \left(\frac{g_{\eta} \pi^2}{30} \right)^{1/4} \rho_r^{3/4},
\qquad
s_T = \frac{4}{3} \left(\frac{g_T \pi^2}{30} \right)^{1/4} \rho_T^{3/4},
\end{equation}
yielding the following entropy ratio:
\begin{equation}
\label{ent1}
\frac{s_T}{s_r} = \alpha^{-3/4}\left(\frac{g_{T}}{g_{\eta}}  \right)^{1/4} \left(\frac{d_{\eta} \sqrt{\pi}  \Lambda_T^{9/2}}{2304 \times 3^{1/4}\sqrt{2}  } \right) \left( \frac{m_{\eta}}{m_{3/2}} \right)^{3/2}.
\end{equation}
To avoid a Polonyi-like problem, we must limit the amount of entropy production $s_T/s_r = \Delta_s$. For a given value of $\Delta_s$, we can derive an upper limit on
$\Lambda_T$ (assuming $g_\eta = g_T$)
\begin{equation}
    \Lambda_T \lesssim 2 \alpha^{1/6} \sqrt{3} \left(\frac{256}{\pi} \right)^{1/9} d_{\eta}^{-2/9} \Delta_s^{2/9} \left( \frac{m_{3/2}}{m_{\eta}} \right)^{1/3}.
    \label{entlim}
\end{equation}
If we consider the representative choices $m_{\eta} \sim 10^{-5} M_P$, $m_{3/2} \sim 10^{-15} M_P$, $d_{\eta} \sim \mathcal{O}(1)$, and $\alpha = 1$, we find the upper bound $\Lambda_T \lesssim 0.003 \Delta_s^{2/9} \sim 0.007$ 
for $\Delta_s \le 100$, thereby mitigating the entropy production problem.

\textbf{II):} {\it Oscillations of the volume modulus $T$ begin after inflaton decay}

We can also consider the case when the damped oscillations of $T$ occur only after the inflaton $\eta$ decays. In this case, we use the Hubble parameter for a radiation-dominated Universe~(\ref{hubble3}) together with the expression $H = \frac{2}{3} \, m_{T}$ to obtain
\begin{equation}
\frac{R_T}{R_{\eta}} \simeq  d_{\eta}^{-1/3} \left(\frac{\alpha}{48}\right)^{1/4} \frac{ m_{\eta}^{1/6} M_P^{1/3}\Lambda_T^{1/2}}{ m_{3/2}^{1/2}} \, .
\label{RTRe}
\end{equation}
Using Eq. (\ref{scale2}) to obtain $R_T/R_{d\eta}$, we can confirm that we are in scenario II when $R_T > R_{d\eta}$,
which occurs when the inequality in Eq. (\ref{lambda1}) is violated. We can now distinguish two cases depending on whether $T$ dominates at the time of its decay, or not.

\textbf{II~a)}: {\it The volume modulus $T$ decays before the Universe becomes dominated by $T$ oscillations}

This case is similar to scenario I b). 
The ratio $R_{dT}/R_\eta$ is given by Eq.~(\ref{Rdtder}),
but the ratio of energies is computed using (\ref{RTRe}) as opposed to (\ref{scaleT}),
giving
\beq
\frac{\rho_T}{\rho_r} = \frac{1}{\alpha} \frac{\sqrt{\pi}}{576\sqrt{6}}  
\frac{M_P \Lambda_T^4}{m_{3/2}} < 1,
\eeq
where the inequality should be satisfied to avoid $T$ domination, and give the limit
\beq
\Lambda_T < 2 \alpha^{1/4} \left( \frac{7776}{\pi} \right)^{1/8} \left( \frac{ m_{3/2}}{M_P}\right)^{1/4}
\label{IIaLT}
\eeq
Note that the limit is independent of $m_\eta$ in this case,
in contrast to the limit in Eq. (\ref{IbLT}). 
The range for $\Lambda_T$ must therefore satisfy the inequality in 
Eq. (\ref{IIaLT}) and violate that in Eq. (\ref{lambda1}).
This is possible if 
\beq
m_{3/2} < \alpha \left( \frac{3}{8\pi} \right)^{1/6} d_\eta^{8/3} \frac{m_\eta^4}{M_P^3},
\label{m32IIa}
\eeq
and corresponds to very small gravitino masses.

\textbf{II~b)}: {\it The volume modulus $T$ decays after the Universe becomes dominated by $T$ oscillations}

This case is similar to I~c). 
The ratio of the entropies is now
\beq
\frac{s_T}{s_r} = \frac{1}{\alpha} \left(\frac{g_{T}}{g_{\eta}}  \right)^{1/4} \left(\frac{\sqrt{\pi}}{1152\sqrt{2}} \right) \left( \frac{\Lambda_T^4 M_P}{m_{3/2}} \right).
\eeq
The limit on $\Lambda_T$ to avoid excessive entropy production is
\beq
\Lambda_T < 2 \alpha^{1/4} \sqrt{3} \left( \frac{128}{\pi} \right)^{1/8} \Delta_s^{1/4} \left( \frac{ m_{3/2}}{M_P} \right)^{1/4}
\label{entlim2}
\eeq
where we are again allowing for entropy increase by a maximum factor of $\Delta_s$.
This case is realized when 
the inequality in (\ref{entlim2}) is satisfied, but that in (\ref{lambda1}) is not. This can occur when 
\beq
m_{3/2} < \alpha \left( \frac{1}{2\pi} \right)^{1/6} d_\eta^{8/3} \frac{m_\eta^4}{M_P^3} \Delta_s^{1/3}.
\label{m32IIb}
\eeq

\subsection{Limits on $\Lambda_T$ from Dark Matter Production}

In this Section we consider the limits on $\Lambda_T$ that avoid the overproduction of the lightest supersymmetric particle (LSP) and related cosmological problems.

We consider first inflaton decay to gravitinos. We calculate the inflaton to two gravitino decay rate using the Lagrangian (\ref{lagsup}), obtaining the decay rate $\Gamma(\phi \rightarrow \psi_{3/2} \psi_{3/2}) \sim \langle G_{\phi} \rangle$. As previously, we assume that $\lambda_2 \ll \lambda_1$. For $\phi$-type models, at the minimum when $\langle T \rangle = \frac{1}{2}$ and $\langle \phi \rangle = 0$, we have $\langle G_{\phi} \rangle = 0$. Similarly, for the case with a twisted inflaton field, we have $\langle G_{\varphi} \rangle = 0$. Therefore, the decays of the inflaton to a pair of gravitinos are negligible, unless there is an additional coupling of $\phi$ to $T$ \cite{EGNO4}.

If the $T$ field decays into a gravitino pair, we have
\begin{equation}
n_{3/2} = 2 \, n_T,
\end{equation}
where $n_{3/2}$ and $n_{T}$ are the number densities of the produced gravitinos and of the decaying $T$ field respectively. Using the following approximations
\begin{equation}
n_{\chi} \simeq n_{3/2}, \qquad s_0 \simeq 7 n_{\gamma}\, ,
\end{equation}
where $n_{\chi}$ is the cold dark matter number density, and $s_0$ is the total entropy density today,  we obtain the following estimate of the cold dark matter abundance ($\rho_c$ is the critical density):
\begin{equation}
\Omega_{\chi} \simeq \frac{7 m_{\chi} n_{3/2} n_{\gamma}}{s_0 \rho_c} \simeq 2.75 \cross 10^{10} h^{-2} \left(\frac{m_{\chi}}{100~\text{GeV}} \right) \left(\frac{n_{3/2}}{s_0} \right) \, .
\end{equation}
Planck 2018 data impose the constraint~\cite{planck18}
\begin{equation}
\Omega_{\chi} h^2 \lesssim 0.12 \, ,
\end{equation}
which leads to the following upper limit on the gravitino-to-entropy ratio:
\begin{equation}
\frac{n_{3/2}}{s_0} \lesssim 4.4 \cross 10^{-12} \left( \frac{100~{\text{GeV}}}{m_{\chi}} \right) \, ,
\label{yieldlimit}
\end{equation}
which is different for scenarios I and II. We first consider gravitinos produced by $T$ decay
and subsequently consider the thermal production of gravitinos which is controlled by the reheat temperature and the value of $d_\eta$.

\textbf{Scenario I:} The number density $n_T = \rho_T/m_T$ depends whether $T$ decays before the inflaton (I a), or after the inflaton but does not dominate the energy density (I b), or dominates the energy density (I c). For (I a), we find:
\begin{equation}
 {\rm (I~a)}: \qquad n_T = \alpha^{7/2} \, \frac{2187\sqrt{3}}{2\pi^2}  \frac{m_{3/2}^5}{M_P^2 \Lambda_T^7} 
     \quad \text{when} \quad R_{dT} < R_{d \eta}, 
\end{equation}    
whereas for (I b) and (I c), we obtain
\begin{equation}
    {\rm (I~b)}:\qquad n_T = \alpha^{11/4} \, \frac{729\times 3^{3/4}}{32\sqrt{2}\pi^{3/2}} d_{\eta}   \frac{m_{3/2}^{7/2} m_{\eta}^{3/2}}{M_P^2 \Lambda_T^{9/2}} \qquad \text{when} \quad R_{dT} > R_{d \eta},~ \text{does not dominate},
\end{equation}
and
\begin{equation}
    {\rm (I~c)}:\qquad n_T = \alpha^{7/2} \, \frac{1}{\sqrt{48}} \left(\frac{1296}{\pi}\right)^2    \frac{m_{3/2}^{5}}{M_P^2 \Lambda_T^{9}} \qquad \text{when} \quad R_{dT} > R_{d \eta},~ \text{ dominates},
    \label{ntIc}
\end{equation}
We obtain the same gravitino-to-entropy ratio in all three cases, which is given by
\begin{equation}
\label{gravyield1}
\frac{n_{3/2}}{s_0} \simeq \sqrt{\alpha} \, \frac{1}{512} \left( \frac{45}{2\pi^2} \right)^{1/4} d_{\eta} g_{\eta}^{-1/4}  \frac{m_{\eta}^{3/2} \Lambda_T^{3}}{m_{3/2} M_P^{1/2}} \, .
\end{equation}
The relevant dark matter yield is given by $n_{\chi}/s \simeq n_{3/2}/s$. In this case, the density parameter $\Omega_{\chi} h^2 \simeq m_{\chi} n_{\chi}/\rho_c$ can be expressed as
\begin{equation}
\label{dmabun1}
\Omega_{\chi} h^2 \simeq 6.6 \times 10^{7} \, \sqrt{\alpha} \, d_{\eta} g_{\eta}^{-1/4} \frac{m_{\eta}^{3/2} \Lambda_T^3}{m_{3/2} M_P^{1/2}} \left(\frac{m_{\chi}}{100~\text{GeV}} \right),
\end{equation}
where we assume that the LSP mass is not much smaller than that of the gravitino.
If we use the nominal value $\Omega_{\chi}h^2 \simeq 0.12$ with~(\ref{dmabun1}), we find
\begin{equation}
    \Lambda_T \lesssim 1.2 \times 10^{-3} \, \alpha^{-1/6} \, d_\eta^{-1/3} g_{\eta}^{1/12} \frac{m_{3/2}^{1/3} M_P^{1/6}}{m_{\eta}^{1/2}}
    \left(\frac{m_{\chi}}{100~\text{GeV}} \right)^{-1/3}.
    \label{DMdensity}
\end{equation}
This is the upper bound on $\Lambda_T$ that is imposed by consistency with the current dark matter density given by the most recent Planck data~\cite{planck18}, making the plausible assumption that the entropy released by the decay of the gravitino is negligible. 

We also consider the thermal production of gravitinos.
Gravitinos are produced during reheating, and the abundance of gravitinos scales with the reheating temperature $T_{\rm RH}$. The gravitino-to-entropy ratio arising from thermal production can be related to the inflaton decay rate, and the hence the inflaton decay coupling $d_\eta$, and is given by~\cite{bbb,egnop}
\begin{equation}
\label{thermyield}
    \frac{n_{3/2}}{s_0} \simeq 2.6\times 10^{-4} d_\eta \left(1 + 0.56 \frac{m_{1/2}^2}{m_{3/2}^2} \right) \left( \frac{m_\eta}{M_P}\right)^{3/2},
\end{equation}
where contributions to the production of transverse modes (1) and longitudinal models (.56 $m_{1/2}^2/m_{3/2}^2$) are included. If we use Eq. (\ref{thermyield}) in
the limit (\ref{yieldlimit}), we obtain an upper limit on the coupling $d_\eta$
\beq
d_\eta < 1.7 \times 10^{-8}\left( \frac{M_P}{m_\eta}\right)^{3/2} \left( \frac{100~{\rm GeV}}{m_\chi} \right) \left(1 + 0.56 \frac{m_{1/2}^2}{m_{3/2}^2} \right)^{-1} \, .
\label{detalimit1}
\eeq
If $m_\eta = 3 \times 10^{13}$~GeV and assuming conservatively the lower limit $m_\chi \gtrsim 100$~GeV
and that $m_{1/2} \ll m_{3/2}$, we obtain
\beq
d_\eta \lesssim 0.4\, .
\label{detalimit}
\eeq
This limit must be respected independently of any assumptions on the stabilization of the volume modulus. 
Note that $d_\eta < 0.4$ corresponds to a limit $y < 2.4 \times 10^{-5}$ where $y$ is a conventionally defined
inflaton coupling, $y = \sqrt{8\pi} d_\eta m_\eta/M_P$,
in agreement with past results \cite{strongmoduli,ENO8,reheating,egnop,egnno1,EGNNO23}. 

Considering the ratio of the yield produced by $T$ decays~(\ref{gravyield1}) to the thermal yield~(\ref{thermyield}), we find
\begin{equation}
    \frac{(n_{3/2}/s_0)_{\rm T \, Decay}}{(n_{3/2}/s_0)_{\rm Thermal}} \simeq 2.4 \, \alpha^{1/2} \left(1 + 0.56 \frac{m_{1/2}^2}{m_{3/2}^2} \right)^{-1} \frac{M_P \Lambda_T^3}{m_{3/2}},
    \label{thermal}
\end{equation}
and thermal production is subdominant when
\begin{equation}
    \Lambda_T \gtrsim 0.75 \, \alpha^{-1/6}
    \left(1 + 0.56 \frac{m_{1/2}^2}{m_{3/2}^2} \right)^{1/3}
   \left( \frac{m_{3/2}}{ M_P} \right)^{1/3} .
\end{equation}

\textbf{Scenario II:} We find the following number density $n_T = \rho_T/m_T$ for case (II a):
\begin{equation}
    {\rm (II~a)}:\qquad n_T \simeq \alpha^{5/2} \frac{2187}{16 \sqrt{2} \pi^{3/2}} \frac{ m_{3/2}^4}{M_P \Lambda_T^5} \quad \text{when} \quad R_{dT} > R_{d \eta},~ \text{does not dominate}.
\end{equation}
For case (II~b), we have the same result as in Eq. (\ref{ntIc}) for $n_T$, but
we find the gravitino-to-entropy ratio to be same for cases II (a,b):
\begin{equation}
    \frac{n_{3/2}}{s_0} \simeq \frac{\alpha^{1/4}}{256} \left(\frac{135}{2\pi^2} \right)^{1/4} g_{\eta}^{-1/4}  \frac{M_P^{1/2} \Lambda_T^{5/2}}{m_{3/2}^{1/2}}.
\end{equation}
In this case, we find the following density parameter
\begin{equation}
    \Omega_{\chi} h^2 \simeq 1.7 \times 10^8 \, \alpha^{1/4} \, g_{\eta}^{-1/4} \frac{M_P^{1/2} \Lambda_T^{5/2}}{m_{3/2}^{1/2}} \left(\frac{m_{\chi}}{100 \, \text{GeV}} \right)
\end{equation}
or
\begin{equation}
    \Lambda_T \lesssim 2.2 \times 10^{-4} \, \alpha^{-1/10} \, g_{\eta}^{1/10}  \left(\frac{m_{3/2}}{M_P} \right)^{1/5} \left(\frac{m_{\chi}}{100 \, \text{GeV}} \right)^{-2/5}
\end{equation}
We next compare this production of gravitinos in moduli decays with thermal production, using the following expression for the gravitino-to-entropy ratio from thermal production
\begin{equation}
    \frac{(n_{3/2}/s_0)_{\rm T \, Decay}}{(n_{3/2}/s_0)_{\rm Thermal}} \simeq 6.2 \, \alpha^{1/4}\, d_{\eta}^{-1} \left(1 + 0.56 \frac{m_{1/2}^2}{m_{3/2}^2} \right)^{-1} \frac{M_P^2 \Lambda_T^{5/2}}{m_{3/2}^{1/2} m_{\eta}^{3/2}},
\end{equation}
and thermal production is subdominant when
\begin{equation}
    \Lambda_T \gtrsim 0.5 \, \alpha^{-1/10} \, d_{\eta}^{2/5} \left(1 + 0.56 \frac{m_{1/2}^2}{m_{3/2}^2} \right)^{2/5} \frac{m_{3/2}^{1/5} m_{\eta}^{3/5}}{M_P^{4/5}} \, .
\end{equation}

Finally, we note that there is a lower limit on $\Lambda_T$ coming from the postulated form of the stabilization terms in Eqs. (\ref{kahuntwist}) and (\ref{kahtwist}). 
Since the stabilization terms in
the K\"ahler potential should be treated as an effective interaction by integrating out fields with masses, $\Lambda_T M_P$, we should require $\Lambda_T M_P > \sqrt{F_T}$~\cite{strongpol,dgmo}, and using~(\ref{fterm}) we find the limit
\begin{equation}
    \Lambda_T >  \alpha^{-1/4} \left(\frac{m_{3/2}}{M_P}\right)^{1/2}
    \label{effint}
\end{equation}
that is imposed by the effective interaction assumption.

\begin{figure}[h!]
\centering
\includegraphics[scale=0.32]{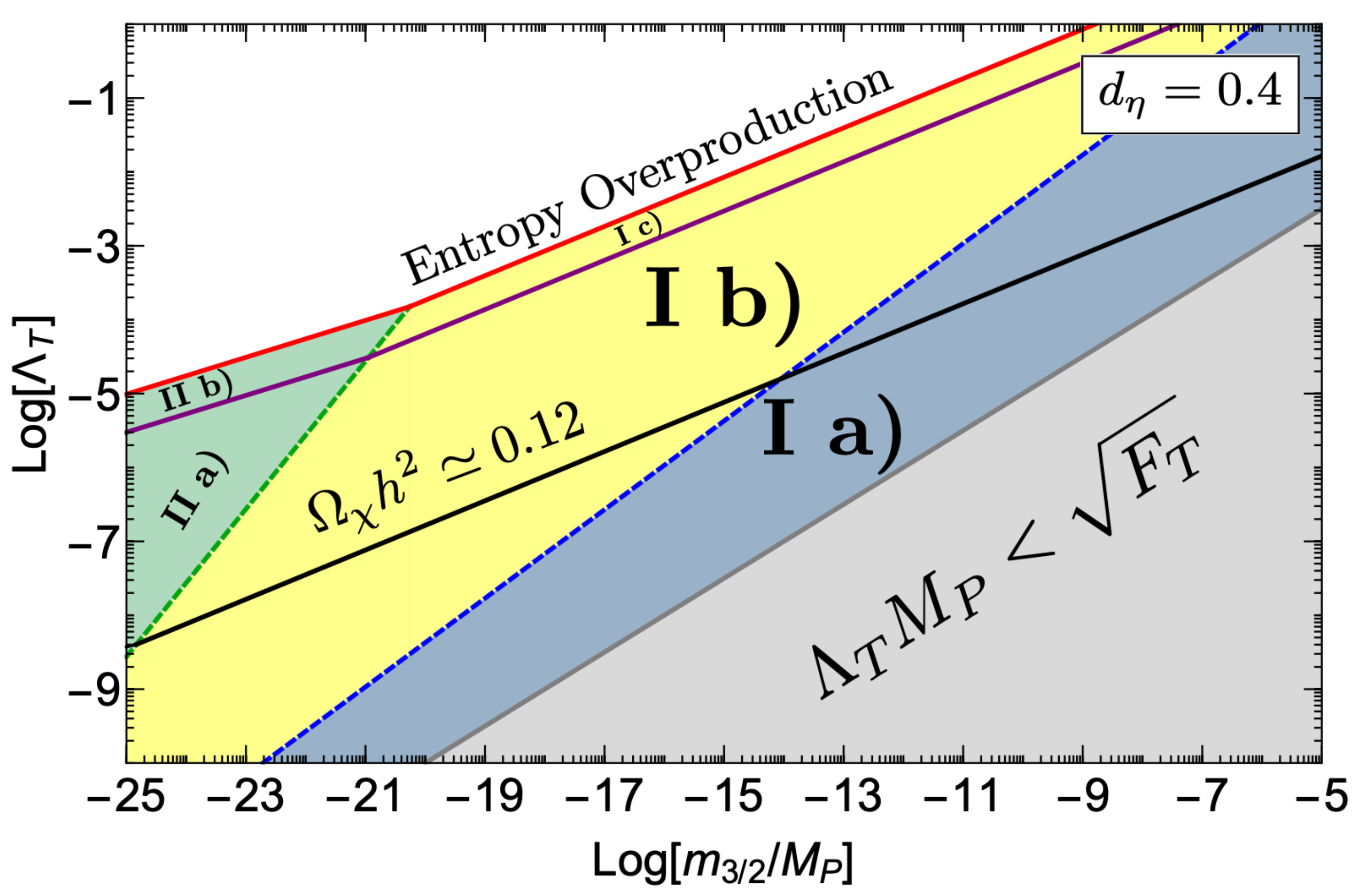}
\includegraphics[scale=0.32]{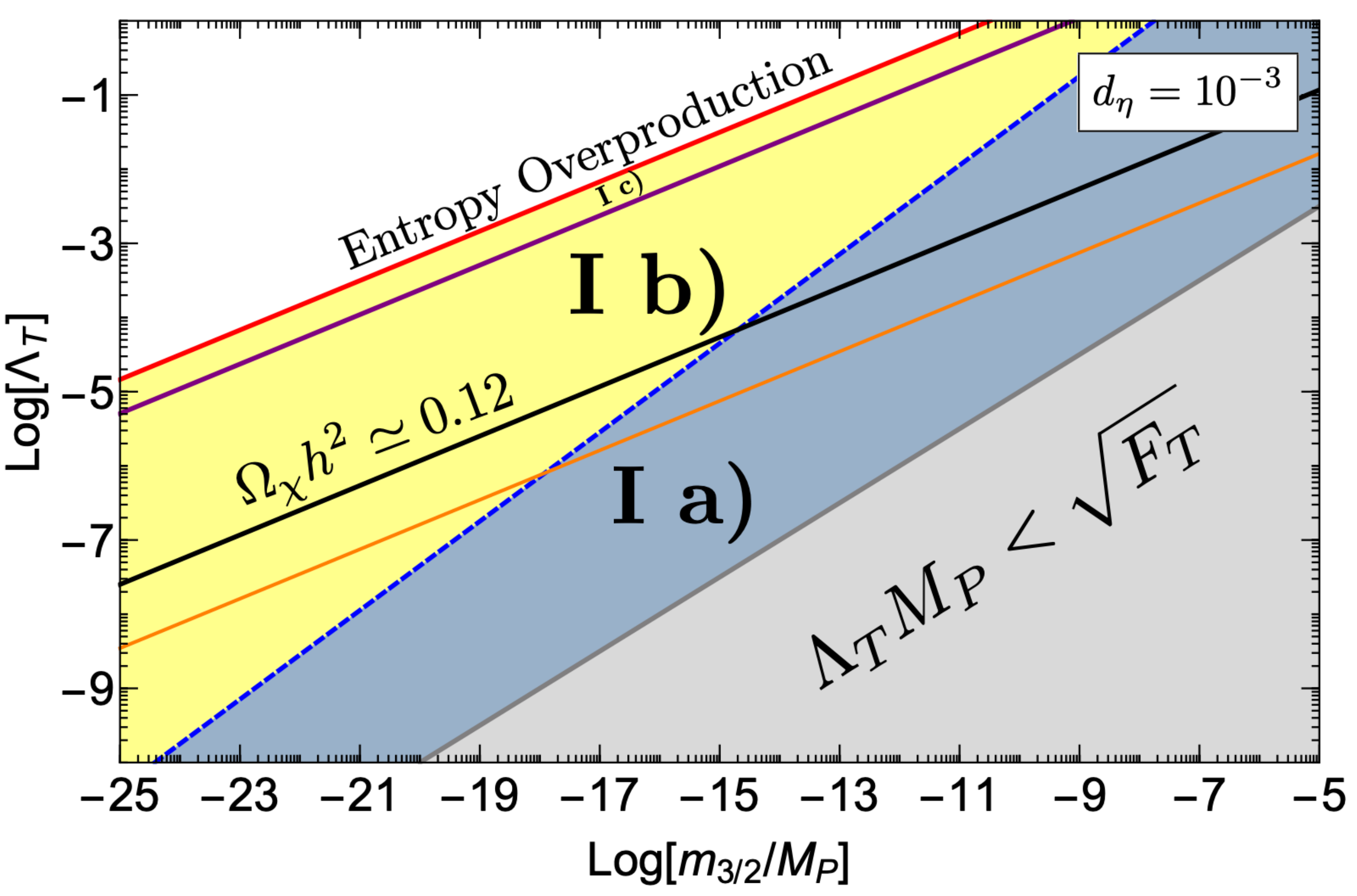}\\
\includegraphics[scale=0.32]{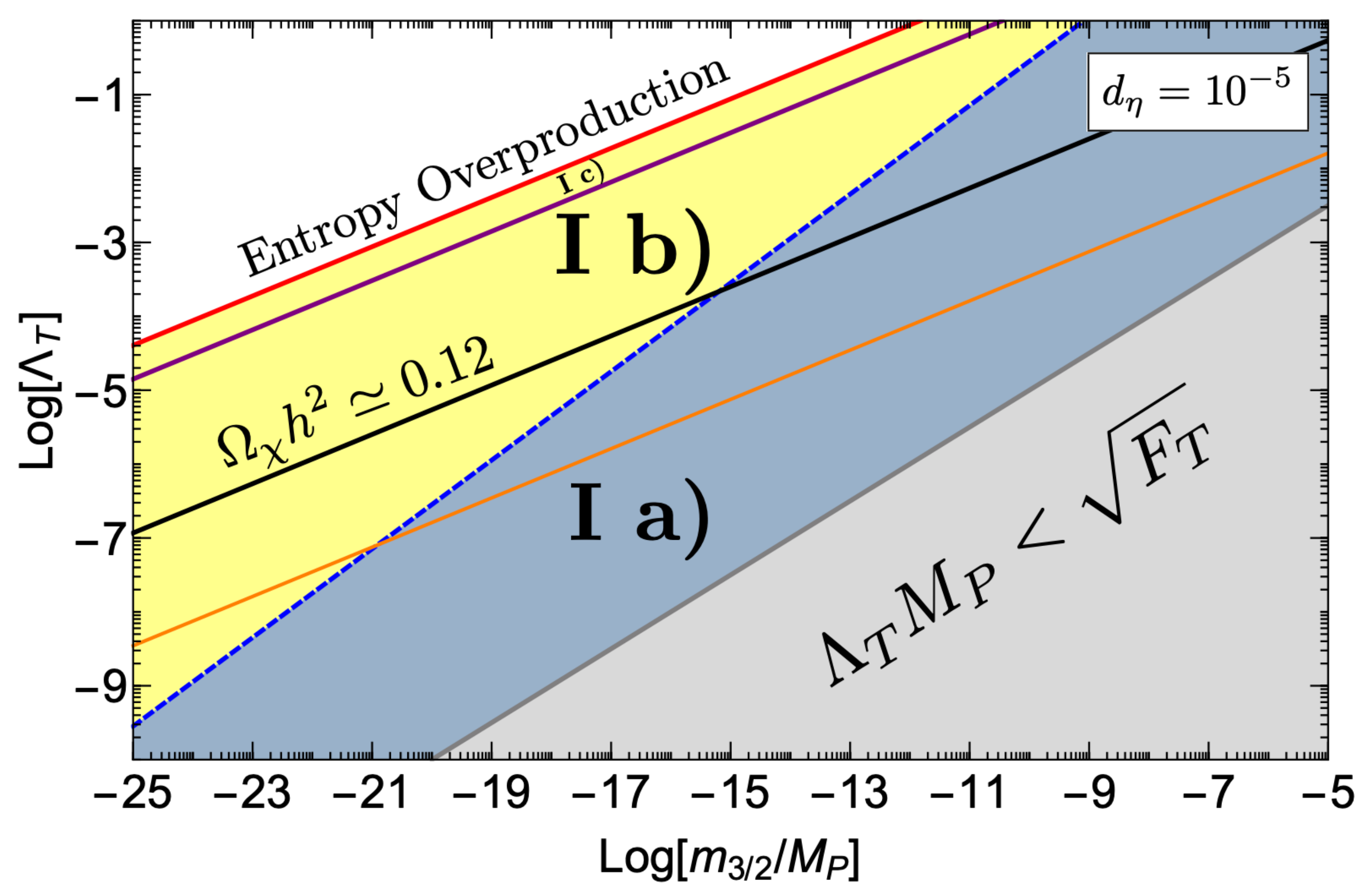}
\caption{\it Plots of the constraints on the modulus stabilization parameter, $\Lambda_T$,
as functions of the gravitino mass, $m_{3/2}$,
for models with $m_\eta = 3 \times 10^{13}$ GeV, $\alpha = 1$ and $d_\eta = 0.4$ (top), $10^{-3}$ (middle)
and $10^{-5}$ (bottom). The regions shaded green, yellow and blue correspond to
Scenarios II~a, b), I~b, c), and I~a), respectively, whereas the grey regions are excluded
by the effective interaction condition (\ref{effint}). Regions between this and the
dark matter density constraint (solid black line) are allowed by all the constraints.}
\label{two}
\end{figure}

We display in Fig.~\ref{two} compilations of the constraints on $\Lambda_T$ as functions of
the gravitino mass for three different values of the coupling $d_\eta$: $0.4$ (which is the largest value
allowed by our analysis - see Eq.(\ref{detalimit})), $10^{-3}$ and $10^{-5}$, 
illustrating their impacts on the various scenarios discussed above. For all our plots we set the inflaton mass equal to its value in the Starobinsky model,  $m_\eta = 3 \times 10^{13}$ GeV, and the curvature parameter $\alpha = 1$. The following are the interpretations of the lines and shadings in the various panels. 
The red lines mark the upper limit on $\Lambda_T$ that is
imposed by the avoidance of entropy overproduction, assuming $\Delta_s \le 100$.
Above the region labelled Scenario II~b) in the top panel, this line is given by Eq. (\ref{entlim2}) and scales as $\Delta_s^{1/4}$, and above the regions labelled Scenario I~c) the red line is determined from Eq.~(\ref{entlim}) and scales as $\Delta_s^{2/9}$. The dashed green line 
in the top panel corresponds to the condition that $T$ oscillations 
begin before inflaton decay - see Eq.~(\ref{lambda1}) - and  
separates Scenarios I) and II).  
Scenario II) is visible only in the upper panel, for large $d_\eta$, and is realized
in the region shaded green between the green dotted line and the entropy overproduction line.
The part of the solid purple line crossing
the green region corresponds to Eq.~(\ref{IIaLT}) and marks 
the boundaries between Scenarios II~a) and II~b) (below and above, respectively). The part of the purple line that crosses
the yellow region corresponds to Eq.~(\ref{IbLT}) and marks the boundaries between Scenarios I~c) and I~b)
(above and below, respectively). The largest value of $m_{3/2}$ in region II~a) is given by Eq.~(\ref{m32IIa})
and that in region II~b) is given by Eq.~(\ref{m32IIb}).
Variants of Scenario I)
are realized in the regions shaded yellow and blue. 
The dashed blue line corresponds to Eq.~(\ref{IaLT}) and marks the boundary between Scenarios I~b) (above)
and I~a) (below): the latter region is shaded blue.
The solid grey line represents the effective interaction condition (\ref{effint}), below
which our parametrization of the dynamics responsible for $T$ stabilization is invalid.

The strongest upper limits on $\Lambda_T$ come from the production of dark matter. 
The thermal production of gravitinos leads to a dark matter abundance which is independent of $\Lambda_T$ (and $m_{3/2}$ when $m_{1/2} \ll m_{3/2}$ - see Eqs.~(\ref{detalimit1})), and depends only on the coupling $d_\eta$. For $d_\eta \simeq 0.4$, thermal production
contributes $\Omega_\chi h^2 =0.12$ everywhere in the plane.
The solid black lines in the three panels show the constraint (\ref{DMdensity}) 
on the contribution to $\Omega_\chi h^2 =0.12$
from $T$ decay, which decreases at lower $\Lambda_T$. At lower values of $d_\eta$, as in the middle and bottom panels of Fig.~\ref{two}, the thermal contribution always 
gives a dark matter density below the Planck limit.  We show as solid orange lines in
these panels, the values of $m_{3/2}, \Lambda_T$ for which the thermal and non-thermal 
contributions are equal.  This line is independent of $d_\eta$ and given by Eq.~(\ref{thermal}).
Below the orange line, the thermal contribution dominates the final dark matter abundance. 
For $d_\eta = 0.4$, as in the upper panel, the orange and black lines coincide. Note that we have fixed $m_\chi = 100$
GeV everywhere in this figure and the relic density scales as $m_\chi$, so that the limit on $\Lambda_T$ scales as $m_\chi^{-1/3}$. Recall also that we have fixed $\alpha = 1$,
though the limit on $\Lambda_T$ scales weakly as $\alpha^{-1/6}$. 

We see  that there are allowed regions in Fig.~\ref{two} below the dark matter density constraint
and above the effective interaction limit. These include regions realized in Scenarios I~a) and I~b),
but not Scenarios I~c) and II). If $m_{3/2} \gtrsim 1$~TeV, the only Scenario that can be realized is I~a),
in which the volume modulus $T$ decays before the inflaton.
However, these restrictions on the possible Scenarios might not apply
if the dark matter density constraint was weakened, e.g., 
if the assumption of R-parity conservation was relaxed.

\section{Conclusions and Prospects}

We have presented in this paper some important phenomenological and cosmological aspects of the no-scale attractor models of inflation that we have introduced previously. These models are based on no-scale supergravity, and include mechanisms for modulus fixing, inflation, supersymmetry breaking and dark energy. As we have discussed, there are models in which inflation is driven by either a modulus field ($T$-type), in which supersymmetry is broken by a Polonyi field,  or a matter field ($\phi$-type) with supersymmetry broken  by  the  modulus  field. We have derived the possible patterns of soft supersymmetry breaking in these different types of models, which depend on the chosen K\"ahler geometries for the matter and inflaton fields, i.e., the parameter $\alpha$ and whether the matter and inflaton are twisted or untwisted. The results are tabulated in Tables~\ref{ttype} and \ref{noscaleTable} for the $T$- and $\phi$-type models, respectively. The patterns of soft supersymmetry breaking found in our analysis include those postulated in the CMSSM, mSUGRA, minimal no-scale supergravity and pure gravity mediation models. Within the framework of no-scale attractor models, phenomenological analyses of the pattern of soft supersymmetry breaking could help to pin down the model type and its K\"ahler geometry. 
We find that there is a direct relation between the scale of supersymmetry breaking and the inflaton mass
in $\phi$-type models.

We have also discussed cosmological constraints on the models from entropy considerations, the density of dark matter, and field stabilization.
These constraints restrict the possible ranges of the quartic parameters in the K\"ahler potential that are used to stabilize the Polonyi field in $T$-type models and the modulus field in the $\phi$-type models. We focus on the $\phi$-type models, in particular, with the results shown in Fig.~\ref{two}. As we see there, the avoidance of entropy and particularly dark matter overproduction require the corresponding stabilization parameter $\Lambda_T$ to be a few orders of magnitude below the Planck scale. We see in Fig.~\ref{two} that there are allowed regions for some of the cosmological scenarios discussed in the text, which could be expanded for small LSP masses or if R-parity is broken.

The key K\"ahler geometry parameter for no-scale attractor models of inflation is the parameter $\alpha$ appearing in Eq.~(\ref{kah2}), which may be related to the form of string compactification. This parameter can be determined by measurements of the CMB observables $r$ and $n_s$, as seen in Eq.~(\ref{observables}).
It is intriguing that this same parameter enters the values of the soft supersymmetry-breaking parameters in Tables~\ref{ttype} and \ref{noscaleTable}, offering the possibility of correlating directly collider and CMB measurements. This is a concrete example how no-scale attractor models could, in the future, serve as bridges between early-Universe cosmology, collider physics and string theory.

\appendix

\section*{A \, Slow-Roll Inflation}
We recall some equations for single-field slow-roll inflation. For a general inflationary potential $V(\phi)$, we find the following Klein-Gordon equation of motion:
\begin{equation}
    \ddot{\phi} = - 3H \dot{\phi} - V'(\phi),
\end{equation}
where the evolution of the scalar field is driven by the potential gradient term $V' = dV/d\phi$.
In order to treat the slow-roll approximation, we introduce the following slow-roll parameters
\begin{equation}
\label{slowroll}
    \epsilon \equiv \frac{1}{2} \left(\frac{V'}{V} \right)^2, \quad \eta \equiv \left( \frac{V''}{V}\right),
\end{equation}
where $\epsilon, |\eta| \ll 1$. 

Next, we introduce the expressions of cosmological observables in terms of the slow-roll parameters. The tensor-to-scalar ratio for single scalar field is given by
\begin{equation}
\label{rpar}
    r \simeq 16 \epsilon,
\end{equation}
and the scalar power spectrum is expressed as
\begin{equation}
\label{nspar}
    n_s - 1 \simeq -6 \epsilon + 2 \eta.
\end{equation}
To solve the flatness and horizon problems, we require the total number of inflationary e-folds to be $N_{*} \simeq 50 - 60$ before the end of inflation. The number of e-folds before the end of inflation is given by
\begin{equation}
\label{efolds}
    N_*(\phi) = \int_{t}^{t_*} H dt \simeq \int_{\phi_{*}}^{\phi} \frac{1}{\sqrt{2 \epsilon}} \, d\phi .
\end{equation}
We find the following expressions for the slow-roll parameters~(\ref{slowroll})
in the $\alpha$-Starobinsky scalar model~(\ref{astaro}):
\begin{equation}
\label{eps1}
    \epsilon = \frac{4}{3 \alpha} \left(1 - e^{\sqrt{\frac{2}{3 \alpha}}x} \right)^{-2},
\end{equation}
and
\begin{equation}
\label{eta1}
   \eta = \frac{4}{3 \alpha} \frac{\left(2 - e^{\sqrt{\frac{2}{3 \alpha}}x} \right)}{\left(1 - e^{\sqrt{\frac{2}{3 \alpha}}x} \right)^2}.
\end{equation}
Combining~(\ref{eps1}) with~(\ref{efolds}), we obtain
\begin{equation}
    N_* = -\frac{3\alpha}{4} \left(1 - e^{\sqrt{\frac{2}{3 \alpha}}x} \right) - \frac{\sqrt{3 \alpha}}{2 \sqrt{2}} x,
\end{equation}
and solving it for $x$, we find
\begin{equation}
\label{fieldx}
    x = \frac{-4 \sqrt{6} N_* - 3 \sqrt{6} \alpha - 3 \sqrt{6}\alpha W_{-1} (-e^{-1-\frac{4N_*}{3\alpha}})}{6\sqrt{\alpha}},
\end{equation}
where $W_k (z)$ is the Lambert $W$ function with $k$ an integer, which is defined as the inverse function of $f(W) = W e^W$ (see also \cite{reheating}). Using the expressions~(\ref{rpar}, \ref{nspar}), the slow-roll parameters~(\ref{eps1}, \ref{eta1}) and expression~(\ref{fieldx}), we find 
\begin{equation}
\label{rpar2}
    r = \frac{64}{3 \alpha} \left(1 + W_{-1} (-e^{-1-\frac{4N_*}{3\alpha}}) \right)^{-2},
\end{equation}
and
\begin{equation}
\label{nspar2}
n_s = 1 - \frac{8}{3 \alpha} \left(\frac{1 - W_{-1} (-e^{-1-\frac{4N_*}{3\alpha}})}{1 + W_{-1} (-e^{-1-\frac{4N_*}{3\alpha}})} \right).
\end{equation}
If we expand~(\ref{rpar2}) and~(\ref{nspar2}) for large~$N_*/\alpha$, we recover~(\ref{cmb}).

\section*{B \, Field Shifts in the Minima}

As noted in Section~\ref{sec:TZX}, the shifts in the minima of the Polonyi field and the inflaton depend on the modular weight of the Polonyi superpotential. Introducing a modular weight $\delta$ for the Polonyi superpotential~(\ref{pol}):
\begin{equation}
 \mu (Z + b) \rightarrow \mu (Z + b) \left(T + \frac{1}{2} \right)^{\delta},
\end{equation}
we find the following shifted VEVs in the untwisted case:

\begin{gather}
\langle T \rangle \simeq \frac{1}{2} + \left(\frac{2 \alpha - 1 - \frac{2 \delta}{3} }{ \alpha c^2}\right) \Delta^2,\\
\langle \phi \rangle \simeq \left( \frac{\sqrt{3}}{c} - \frac{\delta}{\sqrt{3} \alpha c} \right) \Delta,\\
\langle Z \rangle \simeq \frac{\sqrt{\alpha}}{6 \sqrt{3}} \Lambda_Z^2, \\
b  \simeq \frac{1}{\sqrt{3 \alpha}} - \frac{1}{2 \sqrt{3} \alpha^{3/2} c^2}  \left( 1 + 3 \alpha(\alpha - 1) + \frac{\delta(\delta + 3 \alpha(2-6 \alpha + \delta))}{9 \alpha} \right) \Delta^2,
\end{gather}
and in the twisted case:
\begin{gather}
\langle T \rangle \simeq \frac{1}{2} + \left(\frac{2 \alpha}{c^2} - \frac{2 \delta }{3 c^2} \right) \Delta^2,\\
\langle \phi \rangle \simeq \left( \frac{\sqrt{3 \alpha}}{c} - \frac{\delta}{\sqrt{3 \alpha}c} \right) \Delta,\\
\langle Z \rangle \simeq \frac{1}{2 \sqrt{3}} \Lambda_Z^2, \\
b  \simeq \frac{1}{\sqrt{3}} - \left(\frac{(3 \alpha - \delta)^2}{6 \sqrt{3} c^2} \right) \Delta^2.
\end{gather}
These results reduce to those given in the text when $\delta = 0$. 

\newpage
\subsection*{Acknowledgements}

\noindent
The work of JE was supported in part by the United Kingdom STFC Grant ST/P000258/1, and in part by the Estonian Research Council via a Mobilitas Pluss grant. The work of DVN was supported in part by the DOE grant DE-FG02-13ER42020 at Texas A\&M University and in part by the Alexander S. Onassis Public Benefit Foundation. The work of KAO was supported in part by the DOE grant de-sc0011842 at the University of Minnesota.

\end{document}